\input harvmac.tex

\vskip 1.5in
\Title{\vbox{\baselineskip12pt
\hbox to \hsize{\hfill}
\hbox to \hsize{\hfill WITS-CTP-043}}}
{\vbox{
	\centerline{\hbox{Interactions of Massless Higher Spin Fields
		}}\vskip 5pt
        \centerline{\hbox{from String Theory
		}} } }
\centerline{Dimitri Polyakov\footnote{$^\dagger$}
{dimitri.polyakov@wits.ac.za}}
\medskip
\centerline{\it National Institute for Theoretical Physics (NITHeP)}
\centerline{\it  and School of Physics}
\centerline{\it University of the Witwatersrand}
\centerline{\it WITS 2050 Johannesburg, South Africa}
\vskip .3in

\centerline {\bf Abstract}
We construct  vertex operators for massless higher spin
fields in RNS superstring theory and compute some of their
three-point correlators, describing gauge-invariant cubic interactions
of the massless higher spins.
The Fierz-Pauli on-shell conditions 
for the higher spins (including tracelessness and
vanishing divergence) follow from the BRST-invariance conditions
for the vertex operators constructed in this paper.
The gauge symmetries of the massless higher spins emerge as a result
of the BRST nontriviality conditions for these operators, being
equivalent to transformations with the traceless
gauge parameter in the Fronsdal's approach.
The gauge invariance of the interaction terms of the higher spins
is therefore ensured automatically by that of the vertex operators
in string theory. We develop general algorithm to compute
the cubic interactions of the massless higher spins and use it
to explicitly describe  the gauge-invariant interaction 
of two $s=3$ and one $s=4$ massless particles. 

\Date{October 2009}
\vfill\eject
\lref\bianchi{ M. Bianchi, V. Didenko, arXiv:hep-th/0502220}
\lref\sagnottia{A. Sagnotti, E. Sezgin, P. Sundell, hep-th/0501156}
\lref\sorokin{D. Sorokin, AIP Conf. Proc. 767, 172 (2005)}
\lref\fronsdal{C. Fronsdal, Phys. Rev. D18 (1978) 3624}
\lref\coleman{ S. Coleman, J. Mandula, Phys. Rev. 159 (1967) 1251}
\lref\haag{R. Haag, J. Lopuszanski, M. Sohnius, Nucl. Phys B88 (1975)
257}
\lref\weinberg{ S. Weinberg, Phys. Rev. 133(1964) B1049}
\lref\fradkin{E. Fradkin, M. Vasiliev, Phys. Lett. B189 (1987) 89}
\lref\skvortsov{E. Skvortsov, M. Vasiliev, Nucl.Phys.B756:117-147 (2006)}
\lref\skvortsovb{E. Skvortsov, J.Phys.A42:385401 (2009)}
\lref\mva{M. Vasiliev, Phys. Lett. B243 (1990) 378}
\lref\mvb{M. Vasiliev, Int. J. Mod. Phys. D5
(1996) 763}
\lref\mvc{M. Vasiliev, Phys. Lett. B567 (2003) 139}
\lref\brink{A. Bengtsson, I. Bengtsson, L. Brink, Nucl. Phys. B227
 (1983) 31}
\lref\deser{S. Deser, Z. Yang, Class. Quant. Grav 7 (1990) 1491}
\lref\bengt{ A. Bengtsson, I. Bengtsson, N. Linden,
Class. Quant. Grav. 4 (1987) 1333}
\lref\boulanger{ X. Bekaert, N. Boulanger, S. Cnockaert,
J. Math. Phys 46 (2005) 012303}
\lref\metsaev{ R. Metsaev, arXiv:0712.3526}
\lref\siegel{ W. Siegel, B. Zwiebach, Nucl. Phys. B282 (1987) 125}
\lref\siegelb{W. Siegel, Nucl. Phys. B 263 (1986) 93}
\lref\nicolai{ A. Neveu, H. Nicolai, P. West, Nucl. Phys. B264 (1986) 573}
\lref\damour{T. Damour, S. Deser, Ann. Poincare Phys. Theor. 47 (1987) 277}
\lref\sagnottib{D. Francia, A. Sagnotti, Phys. Lett. B53 (2002) 303}
\lref\sagnottic{D. Francia, A. Sagnotti, Class. Quant. Grav.
 20 (2003) S473}
\lref\sagnottid{D. Francia, J. Mourad, A. Sagnotti, Nucl. Phys. B773
(2007) 203}
\lref\labastidaa{ J. Labastida, Nucl. Phys. B322 (1989)}
\lref\labastidab{ J. Labastida, Phys. rev. Lett. 58 (1987) 632}
\lref\mvd{L. Brink, R.Metsaev, M. Vasiliev, Nucl. Phys. B 586 (2000)183}
\lref\klebanov{ I. Klebanov, A. M. Polyakov,
Phys.Lett.B550 (2002) 213-219}
\lref\mve{
X. Bekaert, S. Cnockaert, C. Iazeolla,
M.A. Vasiliev,  IHES-P-04-47, ULB-TH-04-26, ROM2F-04-29, 
FIAN-TD-17-04, Sep 2005 86pp.}
\lref\sagnottie{A. Campoleoni, D. Francia, J. Mourad, A.
 Sagnotti, Nucl. Phys. B815 (2009) 289-367}
\lref\sagnottif{
A. Campoleoni, D. Francia, J. Mourad, A.
 Sagnotti, arXiv:0904.4447}
\lref\sagnottig{D. Francia, A. Sagnotti, J.Phys.Conf.Ser.33:57 (2006)}
\lref\selfa{D. Polyakov, Int.J.Mod.Phys.A20:4001-4020,2005}
\lref\selfb{ D. Polyakov, arXiv:0905.4858}
\lref\selfc{D. Polyakov, arXiv:0906.3663}
\lref\selfd{D. Polyakov, Phys.Rev.D65:084041 (2002)}
\lref\mirian{A. Fotopoulos, M. Tsulaia, Phys.Rev.D76:025014,2007}
\lref\extraa{I. Buchbinder, V. Krykhtin,  arXiv:0707.2181}
\lref\extrab{I. Buchbinder, V. Krykhtin, Phys.Lett.B656:253-264,2007}
\lref\extrac{X. Bekaert, I. Buchbinder, A. Pashnev, M. Tsulaia,
Class.Quant.Grav. 21 (2004) S1457-1464}
\lref \extrad{I. Buchbinder, A. Pashnev, M. Tsulaia,
arXiv:hep-th/0109067}
\lref\extraf{I. Buchbinder, A. Pashnev, M. Tsulaia, 
Phys.Lett.B523:338-346,2001}
\lref\extrag{I. Buchbinder, E. Fradkin, S. Lyakhovich, V. Pershin,
Phys.Lett. B304 (1993) 239-248}
\lref\extrah{I. Buchbinder, A. Fotopoulos, A. Petkou, 
 Phys.Rev.D74:105018,2006}
\lref\bonellia{G. Bonelli, Nucl.Phys.B {669} (2003) 159}
\lref\bonellib{G. Bonelli, JHEP 0311 (2003) 028}
\lref\ouva{C. Aulakh, I. Koh, S. Ouvry, Phys. Lett. 173B (1986) 284}
\lref\ouvab{S. Ouvry, J. Stern, Phys. Lett.  177B (1986) 335}
\lref\ouvac{I. Koh, S. Ouvry, Phys. Lett.  179B (1986) 115 }

\centerline{\bf  1. Introduction}

Constructing the  gauge field theories describing
interacting particles of higher spins (with $s>2$)
is a fascinating and complicated problem that has attracted 
a profound interest over many years since the 30s.
Despite strong efforts by some leading experts in recent
years 
~{\bianchi, \sagnottia, \sorokin, 
\mva, \mvb, \mvc, \deser, \bengt, \siegel, \siegelb,
\nicolai, \damour, \brink, \boulanger,
\labastidaa, \labastidab, \mvd, \mve, \mirian, \extraa,
\extrab, \extrac, \extrad, \extraf, \extrag, \extrah,
\bonellia, \bonellib, \ouva, \ouvab, \ouvac}
 there are still key issues about these theories
that remain unresolved (even for the non-interacting particles;
much more so in the interacting case).
There are several reasons why the higher spin theories
are so complicated. First of all, in order
 to be physically meaningful, these theories need to 
possess
 sufficiently strong gauge symmetries, powerful enough to ensure the absence
of unphysical  (negative norm) states.
For example, in the Fronsdal's description ~{\fronsdal}
the theories  describing
symmetric tensor fields of spin $s$  are invariant under gauge transformations
with the spin $s-1$ traceless parameter.
Theories with the vast gauge symmetries like this are not trivial
to construct even in the non-interacting case,when one
needs to introduce a number of auxiliary fields and
objects like non-local compensators
~{ \sagnottia, \sorokin, \sagnottib, \sagnottic,
\sagnottid}
 Moreover, as the gauge symmetries
in higher spin theories are necessary to eliminate
the unphysical degrees of freedom, 
they must be preserved in the interacting case as well, i.e.
one faces a problem
(even more difficult) of introducing the interactions 
in a gauge-invariant way. In the flat space
things are further complicated 
because of the no-go theorems
( such as Coleman-Mandula theorem ~{\coleman, \haag})
 imposing strong
restrictions on 
conserved charges in interacting theories with a mass gap, 
limiting them to
the scalars and those related to the standard Poincare generators.
Thus Coleman-Mandula theorem  in $d=4$ makes it 
hard to construct consistent interacting theories of higher
spin, at least as long as the locality is preserved,
 despite several examples of higher spin interaction vertices
constructed over the recent years ~{\bengt, \brink, \metsaev}.
In certain cases, such as in AdS backgrounds,  the Coleman-Mandula theorem 
can be bypassed (since there is no well-defined S-matrix in the AdS geometry)
and gauge-invariant interactions  can be  introduced 
consistently - as it has been done in 
 the Fradkin-Vasiliev construction ~{ \mva, \mvb, \mvc, \fradkin,
\skvortsov, \skvortsovb}
The AdS case is particularly interesting since, in the context of the AdS/CFT
correspondence, the higher spin currents in $AdS_4$ have been found to be
dual to the operators in  $d=3$ CFT described by the O(N) model 
~{\klebanov};
also the higher spin dynamics in $AdS_5$ is presumed to be relevant
to the $weakly$ coupled limit of $N=4$ super Yang-Mills theory
in $d=4$.
In non-AdS geometries, however (such as in the flat case),
the no-go theorems do lead to complications, implying, in particular,
that the interacting gauge-invariant theories of higher spins
have to be essentially non-local.

In this paper we approach this problem from the string  theory side
by constructing vertex operators for massless higher spin fields.
It has already been observed some time ago that
string theory  is a particularly effective
and natural  framework to approach the problem of higher spins
~{\sagnottie, \sagnottif, \sagnottig}
at least in the massive case, since the higher spin modes
naturally appear in the massive sector of the theory.
Thus  one  can hope to obtain
the higher field spin theories in the low energy limit
of string theory, by analyzing the worldsheet correlators
of the appropriate vertex operators. 
In  the massless case, discussed in this work, things, however,
are more subtle.
While it is well-known that the massive string modes include higher
spin fields, that can be emitted by the standard vertex operators
(with the standard stringy mass to spin relation),
describing massless higher spin modes in terms of strings is a challenge
since the only vertex  operator in open string theory,
decoupled from superconformal ghost degrees
of freedom (and therefore existing at zero ghost picture) has spin 1. 
Therefore the massless operators for the higher spins are inevitably those
that couple to the worldsheet ghost degrees of freedom and violate the
picture equivalence.
The geometrical reasons for the  existence of such picture-dependent
operators and questions of their BRST invariance and non-triviality
have been discussed in a number of our previous works 
(particularly in ~{\selfb, \selfc}).
In this paper we apply the formalism developed in ~{\selfc, \selfd}
to construct physical vertex operators describing emissions of massless
higher spin fields by an open string. 
We mostly restrict ourselves to totally symmetric higher spin fields,
although it seems to be relatively straightforward to extend the construction,
performed  in this work, to  the higher spins corresponding to
more general Young tableau, as well as
to the case of the multiple families of indices ( e.g. considered in 
~{\sagnottif}).
The BRST-invariance constraints for the vertex operators, 
considered in this paper, lead to the on-shell Pauli-Fierz conditions
for the higher spin fields in space-time, coupled to these operators.
The gauge symmetries of the higher spin fields, on the other hand,
follow from the BRST nontriviality constraints on the appropriate
vertex operators. 
In particular, the BRST nontriviality conditions for massless
symmetric operators of  integer spins from $3$ to $9$,
considered in this work, entail the gauge symmetries equivalent to
those in the Fronsdal's approach (with the tracelessness
condition imposed on the gauge parameter).
Thus the correlation functions of these operators,
computed in this paper, lead to the interaction terms for the higher 
spin fields,
that are gauge-invariant by construction.
The paper is organized as follows. In the Sections 2-5 we  present
the expressions for the vertex operators describing
 emissions of massless symmetric higher spin fields in RNS 
string theory and analyze their BRST invariance and nontriviality 
conditions, leading to the gauge symmetries and the on-shell
conditions for the higher spins fields.
In Section 6  we develop a technique to calculate the 3-point
correlation functions of these operators, particularly using it to derive
the gauge-invariant cubic interaction terms for  $s=3$ and
a $s=4$ higher spins.
In the concluding section we comment on higher order interaction terms
and on the directions for the future work.

\centerline{\bf 2. Vertex Operators for Massless Higher Spins
and BRST Conditions}

We start with presenting the expressions for the vertex operators 
of massless higher spins in RNS superstring formalism.
As was noted above, these operators are essentially coupled to the 
worldsheet ghost fields (in order to ensure the appropriate conformal dimension)
and  violate the equivalence of pictures
(being the elements of nontrivial
 superconformal ghost cohomologies, particularly described in 
~{\selfb, \selfc}).
To compute the their matrix elements, we shall need both negative and
positive ghost picture representations of these operators
(to ensure the ghost anomaly cancellation). 
The expressions for the symmetric massless higher spin operators for
the spin values $3\leq{s}\leq{9}$ at their minimal negative $\phi$-pictures
(i.e. with no local versions at pictures above the minimal one) are given by:

\eqn\grav{\eqalign{
V_{s=3}(p)=H_{a_1a_2a_3}(p)c{e}^{-3\phi}\partial{X^{a_1}}\partial{X^{a_2}}
\psi^{a_3}e^{i{\vec{p}}{\vec{X}}}
\cr
V_{s=4}(p)=H_{a_1...a_4}(p)\eta{e^{-4\phi}}\partial{X^{a_1}}\partial{X^{a^2}}
\partial\psi^{a_3}\psi^{a_4}e^{i{\vec{p}}{\vec{X}}}\cr
V_{s=5}(p)=H_{a_1...a_5}(p){e^{-4\phi}}\partial{X^{a_1}}...\partial{X^{a^3}}
\partial\psi^{a_4}\psi^{a_5}e^{i{\vec{p}}{\vec{X}}}\cr
V_{s=6}(p)
=H_{a_1...a_6}(p)c\eta{e^{-5\phi}}\partial{X^{a_1}}...\partial{X^{a^3}}
\partial^2\psi^{a_4}\partial\psi^{a_5}\psi^{a_6}e^{i{\vec{p}}{\vec{X}}}\cr
V_{s=7}(p)=H_{a_1...a_7}(p)c{e^{-5\phi}}\partial{X^{a_1}}...\partial{X^{a^4}}
\partial^2\psi^{a_5}\partial\psi^{a_6}\psi^{a_7}e^{i{\vec{p}}{\vec{X}}}\cr
V_{s=8}(p)=
H_{a_1...a_8}(p)c\eta{e^{-5\phi}}\partial{X^{a_1}}...\partial{X^{a^7}}
\psi^{a_8}e^{i{\vec{p}}{\vec{X}}}\cr
V_{s=9}(p)=H_{a_1...a_9}(p)c{e^{-5\phi}}\partial{X^{a_1}}...\partial{X^{a^8}}
\psi^{a_9}e^{i{\vec{p}}{\vec{X}}}
}}

where $X^{a}$ and $\psi^a$ are the RNS worldsheet bosons
 and
fermions ($a=0,...,d-1$), 
the ghost fields are bosonized as usual, according to

\eqn\grav{\eqalign{b=e^{-\sigma},c=e^{\sigma}\cr
\gamma=e^{\phi-\chi}\equiv{e^\phi}\eta\cr
\beta=e^{\chi-\phi}\partial\chi\equiv\partial\xi{e^{-\phi}}}}

The vertices for the massless spin fields with $s>9$ can be constructed
similarly, by using the combinations of $\partial{X}$'s
and symmetrized products of $\psi$'s and their derivatives.
Obviously (from simple conformal dimension arguments)
they would have to carry bigger values of minimal negative 
ghost numbers,
 which would make them technically
cumbersome objects  to work with.

For simplicity, in this work  
we shall concentrate on the totally  symmetric
polarization tensors $H_{a_1...a_s}(p)$,
 although it should be relatively straightforward
to generalize the vertices (1) to less symmetric cases
For example, the operators with 2 families of indices 
can be obtained by separating the indices carried
by the derivatives of $X$'s and $\psi$'s into 2 independent
groups.
Let us now turn to the question of the BRST-invariance
and the non-triviality of the vertex operators (1).
We start from the BRST-invariance condition.
For simplicity,  consider the $s=3$ vertex
operator first, all other operators can be analyzed similarly.
For our purposes it is convenient to cast the BRST operator as

\eqn\lowen{Q_{brst}=Q_1+Q_2+Q_3}

where

\eqn\grav{\eqalign{
Q_1=\oint{{dz}\over{2i\pi}}\lbrace{cT-bc\partial{c}}\rbrace\cr
Q_2=-{1\over2}\oint{{dz}\over{2i\pi}}\gamma\psi_a\partial{X^a}\cr
Q_3=-{1\over4}\oint{{dz}\over{2i\pi}}b\gamma^2}}

where T is the full stress-energy tensor.
It is easy to demonstrate that all the vertex operators (1)
commute with $Q_2$ and $Q_3$ of $Q_{brst}$. The commutation with 
$Q_1$, however, requires the constraints on the on-shell fields.
Since all the operators (1) are the  worldsheet
integrals of operators of conformal
dimension 1, they commute with $Q_1$ if the integrands 
are the primary fields, i.e. their OPEs with $T$ don't
contain singularities   stronger than double poles
(along with the on-shell $({\vec{p}})^2=0$ condition). 
Since  $H_{a_1a_2a_3}$ is fully symmetric, the OPE is given by
\eqn\grav{\eqalign{
T(z)\partial{X^{(a_1}}\partial{X^{a_2}}\psi^{a_3)}{e^{i{\vec{p}}\vec{X}}}(w)
H_{a_1a_2a_3}(p)\sim
-{{\eta^{(a_1a_2}\psi^{a_3)}
{e^{i{\vec{p}}\vec{X}}}(w)
H_{a_1a_2a_3}(p)}\over{(z-w)^4}}
\cr
+i{{p^{(a_1}\partial{X^{a_2}}\psi^{a_3)}{e^{i{\vec{p}}\vec{X}}}(w)
H_{a_1a_2a_3}(p)}\over{(z-w)^3}}+O((z-w)^{-2})}}
Therefore the BRST-invariance conditions for the $s=3$ vertex:
\eqn\grav{\eqalign{
H^{a_1}_{a_1a_3}(p)=0\cr
p^{a_1}H_{a_1a_2a_3}(p)=0\cr
p^2H_{a_1a_2a_3}(p)=0}}
are precisely the Pauli-Fierz conditions for the symmetric 
massless higher spins.

Let us now turn to the question of the BRST nontriviality
of the $V_s$ operators  (1).
We look for the conditions to ensure
that $V_s$ cannot be represented as a BRST commutators
with operators in small Hilbert space,i.e. 
for a given $V_s$
there is no operator
$W_s$ such that $V_s=\lbrace{Q_{brst}},W_s\rbrace$.
We start with the operators for massless fields with odd spin values
($s=3,5,7,9$)
that have the following structure if taken at minimal 
 negative ghost pictures $-n$
($n=3$ for $s=3$, $n=4$ for $s=5$ and $n=5$ for $s=7,9$):
\eqn\lowen{
V_s=ce^{-n\phi}F_{{{n^2}\over2}-n+1}(X,\psi)}
where $F_{{{n^2}\over2}-n+1}(X,\psi)$ is the primary matter
field of conformal dimension ${{n^2}\over2}-n+1$
(suppressing all the indices). Then there are only two possible
sources of $W_s$.
The first possibility is that $W_s$
  is proportional to the ghost factor ~$\partial{c}c\partial\xi\partial^2\xi
e^{-(n+2)\phi}$. Then there is a possibility that $V_s$ could be obtained
as a BRST commutator with
\eqn\lowen{W_s=
\partial{c}c\partial\xi\partial^2\xi
e^{-(n+2)\phi}G^{(2n-3)}(\phi,\chi,\sigma)F_{{{n^2}\over2}-n+1}(X,\psi)}
where $G^{(2n-3)}(\phi,\chi,\sigma)$ is the conformal dimension
$2n-3$ polynomial in the derivatives of the bosonized ghost
fields $\phi$, $\chi$ and $\sigma$ that must be chosen
so that
\eqn\lowen{\lbrack{Q_1},W_s\rbrack=0}
Provided that  $G^{(2n-3)}(\phi,\chi,\sigma)$ are chosen to satisfy (9),
it is easy to check that the $W_s$-operators also satisfy
\eqn\grav{\eqalign{
\lbrack{Q_2},W_s\rbrack=0\cr
\lbrack{Q_3},W_s\rbrack=\alpha_nV_s}}
and therefore
\eqn\grav{\eqalign{\lbrack{Q_{brst}},W_s\rbrack=\alpha_nV_s}}
where $\alpha_n$ are the numerical coefficients
that depend on the structure of $G^{(2n-3)}(\phi,\chi,\sigma)$.
A lengthy but straightforward computation shows, however, that
 for all the choices of
 $G^{(2n-3)}(\phi,\chi,\sigma)$, consistent with the condition (9)
 for $n=3,4,5$
 (that are relevant
for the higher spin operators (1) with $3\leq{s}\leq9$) one has
\eqn\grav{\eqalign{\alpha_n=0\cr
n=3,4,5}}
and therefore the higher spin operators cannot be 
 written as commutators of  $Q_{brst}$ with the $W_s$ operators
with the structure (8). The details of the calculation for
$n=3$ case are given in ~{\selfd}; the $n=4$ and $n=5$ cases 
are treated totally similarly, producing
$\alpha_n=0$. At present, we do not know if 
the $\alpha_n$ constants also vanish for $n>5$. This question is 
important in relation with the massless spin operators
with $s>9$.
Thus there are no BRST nontriviality conditions
on the higher spin fields of the $V_s$-operators of the type (7)
due to the $W_s$-operators with the structure (8).
The second, and the only remaining possibility for $V_s$
to be written as BRST commutators
stems from the $W_s$-operators with the ghost structure
$\sim{c}\partial\xi{e^{-(n+1)\phi}}$,satisfying
\eqn\grav{\eqalign{{\lbrack{Q_1},W_s}\rbrack=0\cr
{\lbrack{Q_2},W_s}\rbrack\sim{V_s}\cr
{\lbrack{Q_3},W_s}\rbrack=0}}

The only possible construction for $W_s$ with such a structure is
given by

\eqn\grav{\eqalign{
W_s=c\partial\xi{e^{-(n+1)\phi}}F_{{{n^2}\over2}-n+1}(X,\psi)
(\psi_a\partial{X^a})}}

The operators of this type always commute with $Q_3$ and
produce $V_s$ when commuted with $Q_2$.
Therefore $V_s$ are trivial as long as $W_s$ commute with $Q_1$. 
So $V_s$ are physical operators only if
the commutator 
${\rbrack{Q_1},W_s}\rbrack{\neq}0$, which, in turn, imposes
constraints on the space-time fields and entails the gauge symmetries
for the higher spins.
Let us consider the  particular case of $s=3$, other operators
are analyzed similarly.
The $W_s$-operator of the type (14) for $V_{s=3}$ (1)
is
\eqn\lowen{W_{s=3}(p)=c\partial\xi{e^{-4\phi}}\partial{X^{a_1}}
\partial{X^{a_2}}\psi^{a_3}({\vec{\psi}}{\vec{\partial{X}}})
e^{i{\vec{p}}{\vec{X}}}H_{a_1a_2a_3}(p)}
where, as previously, the $H$ three-tensor  is symmetric
and satisfies the on-shell conditions (6) 
Using the Pauli-Fierz constraints (6) on H, 
one easily finds that $W_{s=3}$ satisfies:
\eqn\grav{\eqalign{
\lbrack{Q_1},W_{s=3}(p)\rbrack=-{i\over2}\partial^2{c}c
\partial\xi{e^{-4\phi}}\partial{X^{a_1}}
\partial{X^{a_2}}\psi^{a_3}({\vec{p}}{\vec{\psi}})
e^{i{\vec{p}}{\vec{X}}}H_{a_1a_2a_3}(p)\cr
\lbrack{Q_2},W_{s=3}(p)\rbrack={d\over2}V_{s=3}(p)\cr
\lbrack{Q_2},W_{s=3}(p)\rbrack=0}}
So the nontriviality of $V_{s=3}$
requires that the right hand side of the commutator 
$\lbrack{Q_1},W_{s=3}(p)\rbrack$ is nonzero.
This leads to the following  nontriviality conditions on the $H$-tensor:
\eqn\lowen{p_{\lbrack{a_4}}H_{a_3\rbrack{a_1}a_2}\neq{0}}
The analysis  of the nontriviality constraints for all other
 odd spin operators (1) ($s=5,7,9$) with  the structure (7)
 is totally similar and 
leads  to  the same conditions on $H_{a_1...a_s}(p)$:
\eqn\lowen{p_{\lbrack{a_{s+1}}}H_{a_s\rbrack{a_1}...a_{s-1}}\neq{0}.}

Next, consider the even spin operators
($s=4,6,8$) that, if taken at their minimal superconformal ghost pictures $-n$
($n=4$ for $s=4$ and $n=5$ for $s=6,8$),
have  the structure 
\eqn\lowen{
V_s=c{\eta}e^{-n\phi}F_{{{n^2}\over2}-n}(X,\psi)}
where $F_{{{n^2}\over2}-n}(X,\psi)$ is again the primary matter
field of conformal dimension ${{n^2}\over2}-n$.
The nontriviality analysis for these operators doesn't differ from
 the odd spin case that we have just described.
As before, there are two potential sources $W_s$
that could imply the triviality of $V_s$, the first is
\eqn\lowen{W_s(p)=\partial{c}c\partial\xi{e^{-(n+2)\phi}}
F_{{{n^2}\over2}-n}(X,\psi)G^{(2n)}(\phi,\chi,\sigma)H(p)}
satisfying (9) - (11)
and 
the second is
\eqn\lowen{W_{s}(p)=c{e^{-n\phi}}
({\vec{\psi}}{\vec{\partial{X}}})
F_{{{n^2}\over2}-n}(X,\psi)G^{(n-1)}(\phi,\chi,\sigma)
e^{i{\vec{p}}{\vec{X}}}H(p)}
satisfying (13)
where, as before, $G^{(h)}(\phi,\chi,\sigma)$ are the conformal
dimension $h$ polynomials in derivatives of the bosonized
ghost fields, chosen so that $W_s$ and $Q_1$ commute.
As previously, lengthy but straightforward analysis (with some help
of Mathematica)
 shows that
all the ghost operators $G^{(2n)}(\phi,\chi,\sigma)$ of (20), leading
to $\lbrack{Q_1},W_s\rbrack=0$ for $W_s$ of the type (20),
 imply $\alpha_n=0$ ($n=4,5$),

implying the nontriviality of $V_s$ without any conditions on
$H(p)$. At the same time, the nontriviality of $V_s$
due to $W_s$ of the type (21) imply the constraints on $H(p)$
identical to (18). Thus the BRST nontriviality constraints
for the massless higher spin operators are summarized by the
condition (18) on $H_{a_1...a_s}(p)$ 
for both  even  and odd values of $s$.
The constraints (18) entail, in turn, the gauge symmetry
transformations for  $H_{a_1...a_s}(p)$ that will be analyzed
in the next section.
\vfill\eject
\centerline{\bf 3. BRST Nontriviality Conditions and Gauge
Symmetries for Higher Spins}

The gauge symmetry for the higher spin fields is the consequence 
of the nontriviality condition (18) for their vertex operators.
It is not difficult to show that the condition (18)
entails the gauge symmetry transformations
\eqn\lowen{H_{a_1...a_n}(p)\rightarrow
{H_{a_1...a_n}(p)+
{p_{(a_1}
\Lambda_{a_2...a_n)}}}}
i.e. the gauge symmetry transformations for a spin $n$
massless field in the Fronsdal's formalism.
 
To show this, consider, for simplicity, the $s=3$ operator, other cases
can be analyzed similarly.
Consider first the case of an arbitrary
(not necessarily symmetric) polarization tensor $H_{a|bc}$
 (which symmetry in $b$ and $c$ is the consequence
of the multiplication by $\partial{X^b}\partial{X^c}$ in the vertex 
operator for $s=3$.

Then the constraint (18) implies  that $H_{a|bc}(p)$
can be shifted by the gauge transformation 
\eqn\lowen{H_{a|bc}(p)\rightarrow
H_{a|bc}(p)+p_{a}\Lambda_{bc}(p)}
provided that the symmetric rank 2 gauge parameter
$\Lambda^{bc}$ is  traceless:
\eqn\lowen{\eta_{bc}\Lambda^{bc}=0}
due to the BRST-invariance conditions (6).
Renaming the indices $a\leftrightarrow{b}$, $a\leftrightarrow{c}$ we get:
\eqn\lowen{H_{b|ac}(p)\rightarrow
H_{b|ac}(p)+p_{b}\Lambda_{ac}(p)}
and
\eqn\lowen{H_{c|ab}(p)\rightarrow
H_{c|ab}(p)+p_{c}\Lambda_{ab}(p)}

Summing together (23), (25), (26) we obtain the transformations
\eqn\lowen{H_{(a|bc)}(p)\rightarrow{H_{(a|bc)}}+
p_{(a}\Lambda_{bc)}(p)}
leading to (22).
Alternatively, one could start with (23), decomposing the left and the
right hand side into two Young diagrams, one fully symmetric
(single row) and another $\Gamma$-like with two rows.
Interestingly, straightforward calculation of the S-matrix
elements involving the vertex operators with the 
double-row polarizations shows them to vanish, so the tensors of
$\Gamma$-like diagrams do not contribute to correlation functions
of the $s=3$ vertex operators (1), with only the symmetric part of
(23) left. 

This concludes the proof that the vertex operators (1)  are the sources
of the massless higher spin fields of spin values $3\leq{s}\leq{9}$
with Pauli-Fierz on-shell conditions and with suitable gauge
symmetries equivalent to those of the Fronsdal's description.
All these properties are consequences of the BRST invariance and nontriviality
conditions for the appropriate vertex operators. Therefore the correlation
functions of these operators, describing the interactions of the massless
higher spins, will by construction lead to the interaction
terms , consistent
with the basic properties of the massless 
 higher spins, including the gauge invariance.
In the following sections we shall 
particularly concentrate on the three-point correlation functions
of the operators (1) leading to the consistent gauge-invariant 
cubic  terms
for interacting massless higher spins. 

\centerline{\bf 4. Vertex Operators for Higher spins:
representations at positive ghost pictures}

Before we start the computation of the correlators
of the higher spin operators, it is necessary
to  obtain their representations in $positive$
ghost pictures, in order to ensure the appropriate ghost number
balance in the correlation functions.

Because the operators (1) violate  picture equivalence,
 higher picture versions cannot be obtained by straightforward
picture-changing transformation (which simply annihilates these operators).
Moreover, there are no local (unintegrated) analogues
of the operators (1) at higher ghost pictures, so all of their
higher picture versions always appear in the integrated form.
 In particular, in this paper we shall need to use,
in addition to unintegrated higher spin
vertex operators (1) at negative ghost pictures $-n-2$
with $n=1,2,3$,
their integrated counterparts at positive ghost pictures $n$.
These counterparts can be constructed by using
the $K$-transformation procedure ~{\selfb, \selfc} which we shall
briefly review below.
Consider one  of unintegrated vertex operators (1) for odd spins
at minimal negative picture $-n-2$ 
(the even spin case is considered analogously).
Such an operator
 has a structure 
\eqn\lowen{V_{-n-2}=ce^{-(n+2)\phi}F_{{{n^2}\over2}+n+1}(X,\psi)}
where, as previously, $F_{{{n^2}\over2}+n+1}(X,\psi)$ the is matter
primary field of conformal dimension ${{n^2}\over2}+n+1$.
Using the fact that the operators $e^{-(n+2)\phi}$
and $e^{n\phi}$ have the same conformal dimension $-{{n^2}\over2}-n$,
one starts with constructing the  charge 
\eqn\lowen{\oint{V_n}\equiv
\oint{{dz}}e^{n\phi}F_{{{n^2}\over2}+n+1}(X,\psi)}
This charge commutes with $Q_1$ since it is a worldsheet integral
of dimension 1 and $b-c$ ghost number zero but doesn't commute with
$Q_2$ and $Q_3$. To make it BRST-invariant, one has to
add the correction terms by using the following procedure 
~{\selfb, \selfc}.
We write
\eqn\grav{\eqalign{\lbrack{Q_{brst}},V_n(z)\rbrack=\partial{U}(z)+W_1(z)
+W_2(z)}}
and therefore
\eqn\lowen{\lbrack{Q_{brst}},
\oint{dz}V_n{\rbrack}=\oint{{dz}}(W_1(z)+W_2(z))}
where 
\eqn\grav{\eqalign{U(z)\equiv{cV_n(z)}\cr
\lbrack{Q_1,V_n}\rbrack=\partial{U}\cr
W_1=\lbrack{Q_2,V_n}\rbrack\cr
W_2=\lbrack{Q_3},V_n\rbrack}}
Introduce the dimension 0 $K$-operator:
\eqn\lowen{K(z)=-4c{e}^{2\chi-2\phi}(z)\equiv{\xi}\Gamma^{-1}(z)}
satisfying
\eqn\lowen{\lbrace{Q_{brst}},K\rbrace=1}
It is easy to check that this operator has a non-singular
operator product with $W_1$:
\eqn\lowen{K(z_1)W_1(z_2)\sim{(z_1-z_2)^{2n}}Y(z_2)+O((z_1-z_2)^{2n+1})}
where $Y$ is some operator of dimension $2n+1$.
Then the complete BRST-invariant operator
can be obtained from $\oint{dz}V_n(z)$
by the following transformation:

\eqn\grav{\eqalign{
\oint{dz}{V_n}(z){\rightarrow}A_n(w)=\oint{dz}V_n(z)+{{1}\over{(2n)!}}
\oint{dz}(z-w)^{2n}:K\partial^{2n}{(W_1+W_2)}:(z)
\cr
+{1\over{{(2n)!}}}\oint{{dz}}\partial_z^{2n+1}{\lbrack}
(z-w)^{2n}{K}(z)\rbrack{K}\lbrace{Q_{brst}},U\rbrace}}
where $w$ is some arbitrary point on the worldsheet.
It is then straightforward to check the invariance
of $A_n$ by using some partial integration along with
the relation (34) as well as the obvious identity
\eqn\lowen{\lbrace{Q_{brst}},W_1(z)+W_2(z)\rbrace=
-\partial(\lbrace{Q_{brst}},U(z)\rbrace)}
Although the invariant operators $A_n(w)$ depend on an
arbitrary point $w$ on the worldsheet, this dependence
is irrelevant in the correlators
 since all the $w$ derivatives  of $A_n$ are BRST exact -
the triviality of the derivatives ensures that
 there will be no $w$-dependence in any correlation
functions involving $A_n$.
Equivalently, 
the positive picture representations $A_n$ (36) for higher spin operators
can also be obtained from minimal negative picture representations
$V_{-n-2}$ by straightforward, but technically more
cumbersome procedure 
by using  the combination of the picture-changing and 
the $Z$-transformation (the analogue of the picture-changing
for the $b-c$-ghosts).

Namely, the $Z$-operator, transforming the $b-c$ pictures (in particular,
mapping integrated vertices to unintegrated)
 given by ~{\selfa}
\eqn\lowen{Z(w)=b\delta(T)(w)=
\oint{dz}(z-w)^3(bT+4c\partial\xi\xi{e^{-2\phi}}T^2)(z)}
where $T$ is the full stress-energy tensor in RNS theory.
The usual picture-changing operator,
transforming the $\beta-\gamma$ ghost pictures, is given by
 $\Gamma(w)=:\delta(\beta)G:(w)
=:e^\phi{G}:(w)$.
Introduce the  $integrated$ picture-changing operators
$R_n(w)$ according to
\eqn\grav{\eqalign{R_{n}(w)=Z(w):\Gamma^{n}:(w)}}
where $:\Gamma^n:$ is the $n$th power of the standard
picture-changing operator:
\eqn\grav{\eqalign{
:\Gamma^{n}:(w)=:e^{{n}\phi}\partial^{n-1}G...\partial{G}G:(w)\cr
\equiv:\partial^{n-1}\delta(\beta)...\partial\delta(\beta)\delta(\beta):}}
Then the positive picture representations for the higher
spin operators $A_n$ can be obtained from the negative ones $V_{-n-2}$ (1)
by the transformation:
\eqn\lowen{A_n(w)=(R_2)^{n+1}(w){V_{-n-2}}(w)}

Since  both $Z$ and $\Gamma$ are BRST-invariant and nontrivial,
the $A_n$-operators by construction 
satisfy the BRST-invariance and non-triviality
conditions identical to those satisfied by their negative picture
counterparts
$V_{-2n-2}$ and therefore lead to the same Pauli-Fierz on-shell
conditions (6) and the gauge symmetries (22), (23) for the higher spin fields.

Below we shall list some concrete examples of the $K$-transformation (36)
applied to the spin $s=3$ and $s=4$ operators that will be used
 in our calculations.
For the $s=3$ operator the above procedure gives
\eqn\grav{\eqalign{V_{s=3}=ce^{-3\phi}\partial{X^{a_1}}\partial{X^{a_2}}
\psi^{a_3}{e^{i{\vec{p}}{\vec{X}}}}H_{a_1a_2a_3}(p)
\rightarrow\oint{dz}V_1\cr
=
H_{a_1a_2a_3}(p)
{\oint}e^{\phi}\partial{X^{a_1}}\partial{X^{a_2}}
\psi^{a_3}{e^{i{\vec{p}}{\vec{X}}}}\cr
\lbrack{Q_1},V_1\rbrack=\partial{U}=
H_{a_1a_2a_3}(p)\partial(c
e^{\phi}\partial{X^{a_1}}\partial{X^{a_2}}
\psi^{a_3}{e^{i{\vec{p}}{\vec{X}}}})\cr
\lbrack{Q_2},V_1\rbrack=W_1={1\over2}H_{a_1a_2a_3}(p)e^{2\phi-\chi}
\lbrace({-}({\vec{\psi}}\partial{\vec{X}})
+i({\vec{p}}{\vec{\psi}})P^{(1)}_{\phi-\chi}+i
({\vec{p}}\partial{\vec{\psi}}))
\partial{X^{a_1}}\partial{X^{a_2}}
\psi^{a_3}{e^{i{\vec{p}}{\vec{X}}}}\cr
+\partial{X^{a_1}}(\partial^2\psi^{a_2}+2\partial\psi^{a_2}P^{(1)}_{\phi-\chi})
\psi^{a_3}-
\partial{X^{a_1}}\partial{X^{a_2}}(\partial^2{X^{a_3}}+
\partial{X^{a_3}}P^{(1)}_{\phi-\chi})\rbrace{e^{i{\vec{p}}{\vec{X}}}}
\cr
\lbrack{Q_3},V_1\rbrack=W_2=-{1\over4}H_{a_1a_2a_3}(p)e^{3\phi-2\chi}
P^{(1)}_{2\phi-2\chi-\sigma}
\partial{X^{a_1}}\partial{X^{a_2}}
\psi^{a_3}{e^{i{\vec{p}}{\vec{X}}}}}}
where the conformal weight $n$ polynomials in the derivatives
of the ghost fields $\phi,\chi, \sigma$ are defined according
to ~{\selfb, \selfc}:

\eqn\lowen{P^{(n)}_{f(\phi,\chi,\sigma)}=e^{-f(\phi(z),\chi(z),\sigma(z))}
{{\partial^{n}}\over{\partial{z^n}}}e^{f(\phi(z),\chi(z),\sigma(z))}}
where $f$ is some linear function in $\phi,\chi,\sigma$.
For example, $P^{(1)}_{\phi-\chi}=\partial\phi-\partial\chi$, etc.
Note that the  product (43) is defined in the algebraic sense 
(not as an operator product).

Accordingly,
\eqn\grav{\eqalign{:K\partial^2{W_1}:
=4H_{a_1a_2a_3}(p)c\xi
\lbrace({-}({\vec{\psi}}\partial{\vec{X}})
+i({\vec{p}}{\vec{\psi}})P^{(1)}_{\phi-\chi}+i
({\vec{p}}\partial{\vec{\psi}}))
\partial{X^{a_1}}\partial{X^{a_2}}
\psi^{a_3}{e^{i{\vec{p}}{\vec{X}}}}\cr
+\partial{X^{a_1}}(\partial^2\psi^{a_2}+2\partial\psi^{a_2}P^{(1)}_{\phi-\chi})
\psi^{a_3}-
\partial{X^{a_1}}\partial{X^{a_2}}(\partial^2{X^{a_3}}+
\partial{X^{a_3}}P^{(1)}_{\phi-\chi})\rbrace{e^{i{\vec{p}}{\vec{X}}}}\cr
:K\partial^2W_2:=H_{a_1a_2a_3}(p){\lbrace}-\partial^2(
e^{\phi}\partial{X^{a_1}}\partial{X^{a_2}}
\psi^{a_3}{e^{i{\vec{p}}{\vec{X}}}})+
P^{(2)}_{2\phi-2\chi-\sigma}
e^{\phi}\partial{X^{a_1}}\partial{X^{a_2}}
\psi^{a_3}{e^{i{\vec{p}}{\vec{X}}}}\rbrace}}
and
\eqn\grav{\eqalign{
:\partial^{2n+1}K{K}\lbrace{Q_{brst}},U\rbrace:=
-24H_{a_1a_2a_3}(p)\partial{c}c\partial\xi\xi{e^{-\phi}}
\partial{X^{a_1}}\partial{X^{a_2}}
\psi^{a_3}{e^{i{\vec{p}}{\vec{X}}}}\cr
:\partial^{m}K{K}\lbrace{Q_{brst}},U\rbrace:=0 (m<2n+1)}}
and therefore, upon integrating out total derivatives,
 the  complete BRST-invariant expression
for the $s=3$ operator at picture 1 is
\eqn\grav{\eqalign{A_{s=3}(w)=
H_{a_1a_2a_3}(p)\oint{dz}(z-w)^2\lbrace
{1\over2}
P^{(2)}_{2\phi-2\chi-\sigma}
e^{\phi}\partial{X^{a_1}}\partial{X^{a_2}}
\psi^{a_3}
\cr+
2c\xi
\lbrack({-}({\vec{\psi}}\partial{\vec{X}})
+i({\vec{p}}{\vec{\psi}})P^{(1)}_{\phi-\chi}+i
({\vec{p}}\partial{\vec{\psi}}))
\partial{X^{a_1}}\partial{X^{a_2}}
\psi^{a_3}{e^{i{\vec{p}}{\vec{X}}}}\cr
+\partial{X^{a_1}}(\partial^2\psi^{a_2}+2\partial\psi^{a_2}P^{(1)}_{\phi-\chi})
\psi^{a_3}-
\partial{X^{a_1}}\partial{X^{a_2}}(\partial^2{X^{a_3}}+
\partial{X^{a_3}}P^{(1)}_{\phi-\chi})\rbrack
\cr
-12\partial{c}c\partial\xi\xi{e^{-\phi}}
\partial{X^{a_1}}\partial{X^{a_2}}
\psi^{a_3}\rbrace{e^{i{\vec{p}}{\vec{X}}}}}}

To abbreviate notations for our calculations of the correlation
functions in the following sections, it is convenient
to write the vertex operator $A_{s=3}$ (46) as a sum 

\eqn\lowen{A_{s=3}=A_0+A_1+A_2+A_3+A_4+A_5+A_6}
where
\eqn\lowen{A_0(w)={1\over2}H_{a_1a_2a_3}(p)\oint{dz}(z-w)^2
P^{(2)}_{2\phi-2\chi-\sigma}
e^{\phi}\partial{X^{a_1}}\partial{X^{a_2}}
\psi^{a_3}{e^{i{\vec{p}}{\vec{X}}}}(z)}
and
\eqn\lowen{A_6(w)=-12H_{a_1a_2a_3}(p)\oint{dz}(z-w)^2
\partial{c}c\partial\xi\xi{e^{-\phi}}
\partial{X^{a_1}}\partial{X^{a_2}}
\psi^{a_3}\rbrace{e^{i{\vec{p}}{\vec{X}}}}(z)}
have ghost factors proportional to
$e^\phi$ and $\partial{c}c\partial\xi\xi{e^{-\phi}}$ respectively
and the rest of the terms carry ghost factor proportional to
$c\xi$:
\eqn\grav{\eqalign{A_1(w)=-2H_{a_1a_2a_3}(p)\oint{dz}(z-w)^2
c\xi({\vec{\psi}}\partial{\vec{X}})
\partial{X^{a_1}}\partial{X^{a_2}}
\psi^{a_3}
{e^{i{\vec{p}}{\vec{X}}}}(z)\cr
A_2(w)=2iH_{a_1a_2a_3}(p)\oint{dz}(z-w)^2
c\xi({\vec{p}}{\vec{\psi}})P^{(1)}_{\phi-\chi}
\partial{X^{a_1}}\partial{X^{a_2}}
\psi^{a_3}
{e^{i{\vec{p}}{\vec{X}}}}(z)\cr
A_3(w)=
2iH_{a_1a_2a_3}(p)\oint{dz}(z-w)^2
c\xi({\vec{p}}\partial{\vec{\psi}})
\partial{X^{a_1}}\partial{X^{a_2}}
\psi^{a_3}
{e^{i{\vec{p}}{\vec{X}}}}(z)\cr
A_4(w)=
2H_{a_1a_2a_3}(p)\oint{dz}(z-w)^2
c\xi(\partial^2\psi^{a_2}+2\partial\psi^{a_2}P^{(1)}_{\phi-\chi})
\psi^{a_3}
{e^{i{\vec{p}}{\vec{X}}}}(z)\cr
A_5(w)=
-2H_{a_1a_2a_3}(p)\oint{dz}(z-w)^2c\xi
\partial{X^{a_1}}\partial{X^{a_2}}(\partial^2{X^{a_3}}+
\partial{X^{a_3}}P^{(1)}_{\phi-\chi})
{e^{i{\vec{p}}{\vec{X}}}}(z)}}

Analogously, the $K$-operator procedure applied to the $s=4$
vertex operator in (1) leads to the positive picture representation
of the $s=4$ operator given by
\eqn\lowen{B_{s=4}=B_0+B_1+B_2+B_3+B_4+B_5+B_6}

where
\eqn\lowen{B_0(w)={1\over{2}}H_{a_1a_2a_3a_4}(p)\oint{dz}(z-w)^2
P^{(2)}_{2\phi-2\chi-\sigma}{\eta}e^{2\phi}\partial{X^{a_1}}
\partial{X^{a_2}}\partial\psi^{a_3}\psi^{a_4}{e^{i{\vec{p}}{\vec{X}}}}(z)}
and
\eqn\lowen{{B_7}(w)=-12
H_{a_1a_2a_3a_4}(p)\oint{dz}(z-w)^2
\partial{c}c\xi\partial{X^{a_1}}
\partial{X^{a_2}}\partial\psi^{a_3}\psi^{a_4}{e^{i{\vec{p}}{\vec{X}}}}(z)}
carry the ghost factors $\sim\eta{e^{2\phi}}$ and $\sim\partial{c}{c}\xi$
respectively, while the rest of the terms carry the ghost factor
$\sim{c}e^\phi$:
\eqn\grav{\eqalign{
{B_1}(w)=-2
H_{a_1a_2a_3a_4}(p)\oint{dz}(z-w)^2
c{e}^{\phi}({\vec{\psi}}\partial{\vec{X}})
\partial{X^{a_1}}
\partial{X^{a_2}}\partial\psi^{a_3}\psi^{a_4}{e^{i{\vec{p}}{\vec{X}}}}(z)\cr
{B_2}(w)=2i
H_{a_1a_2a_3a_4}(p)\oint{dz}(z-w)^2
c{e}^{\phi}({\vec{p}}\partial{\vec{\psi}})P^{(1)}_{\phi-\chi}
\partial{X^{a_1}}
\partial{X^{a_2}}\partial\psi^{a_3}\psi^{a_4}{e^{i{\vec{p}}{\vec{X}}}}(z)\cr
{B_3}(w)=2i
H_{a_1a_2a_3a_4}(p)\oint{dz}(z-w)^2
c{e}^{\phi}({\vec{p}}\partial{\vec{\psi}})
\partial{X^{a_1}}
\partial{X^{a_2}}\partial\psi^{a_3}\psi^{a_4}{e^{i{\vec{p}}{\vec{X}}}}(z)\cr
B_4(w)=
2H_{a_1a_2a_3a_4}(p)\oint{dz}(z-w)^2
P^{(2)}_{\phi-\chi}ce^\phi\partial{X^{a_1}}
\partial^2\psi^{a_2}\partial\psi^{a_3}\psi^{a_4}
{e^{i{\vec{p}}{\vec{X}}}}(z)\cr
B_5(w)=
2H_{a_1a_2a_3a_4}(p)\oint{dz}(z-w)^2{c}e^\phi
\partial{X^{a_1}}\partial{X^{a_2}}({1\over2}\partial^3{X^{a_3}}
\cr
+\partial^2{X^{a_3}}P^{(1)}_{\phi-\chi}+{1\over2}
\partial{X^{a_3}}P^{(2)}_{\phi-\chi})\psi^{a_4}
{e^{i{\vec{p}}{\vec{X}}}}(z)\cr
B_6(w)=
-2H_{a_1a_2a_3a_4}(p)\oint{dz}(z-w)^2{c}e^\phi
\partial{X^{a_1}}\partial{X^{a_2}}(\partial^2{X^{a_3}}
+\partial{X^{a_3}}P^{(1)}_{\phi-\chi})\partial\psi^{a_4}
{e^{i{\vec{p}}{\vec{X}}}}(z)}}
The procedure is totally similar for the operators
in (1) with $s\geq{5}$ which positive picture representations
can be constructed analogously;
however, higher ghost number operators generally consist of
bigger number of terms, so the manifest expressions
for operators with higher $n$ become quite cumbersome.

\centerline{\bf 5. $\xi$-dependence of Higher Spin Vertices:
a comment 
}

One  property of the higher spin vertices
which may seem unusual
is their manifest dependence on the zero mode of $\xi$
in positive picture representations which poses a question
whether the states created by these
 operators are outside the small Hilbert space.
It is not difficult to see, however, that
the manifest $\xi$-dependence of the operators (46)-(54) is just
the matter of the gauge and can be removed by  suitable 
picture-changing transformation. Indeed, it is straightforward 
to show that  all the operators with the structure (36)
can be represented as BRST commutators in the $large$
Hilbert space:
\eqn\lowen{A_n=const\times\lbrace{Q_{brst}},
\oint{dz}(z-w)^{2n}\xi\partial\xi{c}e^{(n-2)\phi}
F_{{{n^2}\over2}+n+1}(X,\psi)\rbrace}
with the matter operators $F_{{{n^2}\over2}+n+1}(X,\psi)$ taken
the same as in (1).
But since $\lbrace{Q_{brst}},\xi\rbrace=\Gamma$ is
 a picture-changing operator, the expression (55)
is actually given by picture-changing transformation
of an operator $inside$ the small Hilbert space.
In fact, the $\xi$-dependence
of the $A_n$-vertices is inherited from
the structure of the $Z$-operator in the map (41).
 The $Z$-operator (38) relating the $b-c$ ghost pictures
also manifestly depends on $\xi$ but,
as one can cast it as a BRST commutator in the large
Hilbert space  ~{\selfc}: 
$$Z(w)=16\lbrace{Q_{brst}},\oint{dz}(z-w)^3
bc\partial\xi\xi{e^{-2\phi}T}\rbrace$$
such a dependence is also the matter of the gauge,
in analogy to the case explained above.

\centerline{\bf 6. Three-Point Correlation Functions
and Cubic Interactions of Higher Spin Fields}

In this section we derive the gauge-invariant
cubic interaction terms for the higher spinors
by computing the three-point functions of the vertex
operators (1), (36), (46), (51). The gauge invariance is 
the consequence of the BRST non-triviality conditions for the 
vertex operators and is thus
ensured by construction. For simplicity, we shall particularly concentrate
on derivation of the cubic interaction
of $s=3$ and $s=4$ spin fields - mainly because the vertex operators
for these fields have relatively simple structure;
however the calculation performed in this section is straightforward
to generalize to cases of other massless integer higher spins.
Before we start, it is useful to introduce an
object that we shall refer to as ``interaction block'' 
$T_{p,q,r|s}^{a_1...a_p|b_1...b_q|c_1...c_r}(p_1,p_2,p_3)$
and that will
play an important  role in our calculations.
Consider a three-point correlation function
\eqn\grav{\eqalign{
A^{a_1...a_p|b_1...b_q|c_1...c_r}(z,w,u;p_1,p_2,p_3)
=<V_1(z,p_1)V_2(w,p_2)V_3(u,p_3)>\cr
=<\partial{X^{a_1}}...\partial{X^{a_p}}e^{i{\vec{p}}_1{\vec{X}}}(z)
\partial{X^{b_1}}...\partial{X^{b_q}}e^{i{\vec{p}}_2{\vec{X}}}(w)
\partial{X^{c_1}}...\partial{X^{c_r}}e^{i{\vec{p}}_3{\vec{X}}}(u)>}}
with the momenta ${\vec{p}}_1,{\vec{p}}_2,{\vec{p}}_3$
satisfying $p_1^2=p_2^2=p_3^2=0$, so the operators
have conformal dimensions $p$,$q$ and $r$ respectively.
Take the limit $u\rightarrow{\infty}$ in which 
$A^{a_1...a_p|b_1...b_q|c_1...c_r}$  becomes function of $u$ and $z-w$
It is not difficult  to see that it will consist of terms
which asymptotic behaviour $u^{-s}$ will
range from $s=r$ to $s=p+q+r$, depending on pairing arrangements
of $\partial{X}$'s . Then  the interaction blocks
$T_{p,q,r|s}^{a_1...a_p|b_1...b_q|c_1...c_r}(p_1,p_2,p_3)$ are defined
as the coefficients in the expansion:
\eqn\grav{\eqalign{lim_{u\rightarrow\infty}
A^{a_1...a_p|b_1...b_q|c_1...c_r}(z,w,u;p_1,p_2,p_3)
\cr
=(z-w)^{{\vec{p}}_1{\vec{p}}_2}\sum_{s=r}^{p+q+r}
{{T_{p,q,r|s}^{a_1...a_p|b_1...b_q|c_1...c_r}(p_1,p_2,p_3)}\over
{u^s(z-w)^{p+q+r-s}}}}}
It is not difficult  to obtain
the manifest expressions.
for $T_{p,q,r|s}^{a_1...a_p|b_1...b_q|c_1...c_r}(p_1,p_2,p_3)$
Consider the contribution
defined by 
$n_1$ pairings between $\partial{X}$'s of $V_1(z)$ and 
those of $V_3(u)$; $n_2$ pairings
 between $\partial{X}$'s of $V_2(w)$ and 
those of $V_3(u)$ and $n_3$ pairings  between $\partial{X}$'s of $V_1(z)$ and 
those of $V_2(w)$. In addition,
let this contribution be characterized by
the numbers $m_1,...,m_6$ where
$m_1$ and $m_2$ are the numbers of pairings of
$\partial{X}$'s in $V_3(u)$ with the exponents
$e^{i{\vec{p}}_1{\vec{X}}}(z)$ and $e^{i{\vec{p}}_2{\vec{X}}}(w)$
respectively;  $m_3$ and $m_4$ are the numbers of pairings
of $\partial{X}$'s in $V_1(z)$ with the exponents
$e^{i{\vec{p}}_2{\vec{X}}}(w)$ and $e^{i{\vec{p}}_3{\vec{X}}}(u)$
respectively and, finally,
$m_5$ and $m_6$
 are the numbers of pairings
of $\partial{X}$'s in $V_2(w)$ with the exponents
$e^{i{\vec{p}}_1{\vec{X}}}(z)$ and $e^{i{\vec{p}}_3{\vec{X}}}(u)$.
It is not difficult to see that 
$T_{p,q,r|s}^{a_1...a_p|b_1...b_q|c_1...c_r}(p_1,p_2,p_3)$
is given by the sum of the diagrams with each diagram
completely characterized by the set $\lbrace{n_i}\rbrace,\lbrace{m_j},\rbrace$
($i=1,2,3;j=1,...,6$) with the following constraints on
$n_i$ and $m_j$, defined by the number of $\partial{X}$'s
 each of the vertices ($p,q$ and $r$)
, as well as by the $u$-asymptotics, given by $s$ 
\eqn\grav{\eqalign{
n_1+n_2+m_1+m_2=r\cr
n_1+n_3+m_3+m_4=p\cr
n_2+n_3+m_5+m_6=q\cr
2n_1+2n_2+m_1+m_2+m_4+m_6=s\cr
0\leq{m_1,m_2}\leq{r}\cr
0\leq{m_3,m_4}\leq{p}\cr
0\leq{m_5,m_6}\leq{q}\cr
0\leq{n_1}\leq{min}(p,r)\cr
0\leq{n_2}\leq{min}(q,r)\cr
0\leq{n_3}\leq{min}(p,q)}}
The symmetry factor for each diagram is easily calculated
to give
\eqn\lowen{N_{symm}={{p!q!r!}\over{{\prod_{i=1}^3}n_i!{\prod_{j=1}^6}m_j!}}}
Using the operator products:
\eqn\grav{\eqalign{
\partial{X^a}(z)\partial{X^b}(w)\sim{-}{{\eta^{ab}}\over{(z-w)^2}}\cr
\partial{X^a}(z)e^{i{\vec{p}}{\vec{X}}}(w)
\sim{{-ip^a{e^{i{\vec{p}}{\vec{X}}}(w)}}\over{z-w}}
}}
it is straightforward to find
\eqn\grav{\eqalign{
T_{p,q,r|s}^{a_1...a_p|b_1...b_q|c_1...c_r}(p_1,p_2,p_3)
\cr
=\sum_{{\lbrace{m}\rbrace},{\lbrace{n}\rbrace}}
{{(-1)^{n_1+n_2+n_3+m_1+m_2+m_4}}\over{\prod_{i=1}^3{n_i}!\prod_{j=1}^6{m_j}!}}
\prod_{k=1}^{n_1}\eta^{a_kc_k}\prod_{k=1}^{n_3}\eta^{a_{n_1+k}b_k}
\prod_{k=1}^{n_2}\eta^{b_{n_3+k}c_{n_2+k}}\cr
\prod_{k=1}^{m_1}(ip_1)^{c_{n_1+n_2+k}}\prod_{k=1}^{m_2}
(ip_2)^{c_{n_1+n_2+m_1+k}}\prod_{k=1}^{m_3}(ip_2)^{a_{n_3+n_1+k}}
\cr
\prod_{k=1}^{m_4}(ip_2)^{a_{n_3+n_1+m_3+k}}
\prod_{k=1}^{m_5}(ip_1)^{b_{n_2+n_3+k}}
\prod_{k=1}^{m_6}(ip_3)^{b_{n_2+n_3+m_5+k}}\cr
+Symm\lbrace(a_1,...,a_p);(b_1,...,b_q);(c_1,...,c_r)\rbrace
}}
where the sum over $n_i$ and $m_j$ is taken over all the values
satisfying (58) and the symmetrization is performed
within each family of indices
$(a_1,...,a_p)$, $(b_1,...,b_q)$ and $(c_1,...,c_r)$
(note that this symmetrization  absorbs the factor of $p!q!r!$
in the numerator of (59))
Note that, in the particular case of $s=2$ the blocks (57), (61)
simply define the 3-point correlators of $massive$
fully symmetric higher spin particles in bosonic
string theory, with the spins
$p,q$ and $r$ respectively and with the square masses
$p-1$, $q-1$ and $r-1$ respectively.
Unlike the massless case, these massive interactions
are not in conflict with the no-go theorems and
are described by the standard massive vertex operators with 
the elementary ghost structure.
Having obtained the expressions for the interaction blocks
(57), (61), we are now prepared to proceed
with the computation of the three-point correlators of the  vertex operators
(1), (36) that determine the gauge-invariant cubic interactions of 
massless higher spins.
As was noted above, we shall concentrate
on the three-point correlation function of two $s=3$ and one $s=4$
operators; other three-point functions
of higher spin operators (1), (36) can be obtained in a similar way.
In order to ensure the cancellation of
 all the ghost number anomalies, the correct ghost number
balance requires that each correlation function has total
$b-c$ ghost number equal to 3, superconformal $\phi$-ghost number equal to $-2$
and superconformal $\chi$-ghost number equal to 1.
This means that in the three-point correlation function
$<V_{s=3}V_{s=3}V_{s=4}>$ two operators must be taken in the
positive picture representation (46), (51) and one at the negative
picture (1). It is particularly convenient
to choose $V_{s=4}$ and one of $V_{s=3}$ at positive pictures and the
remaining $V_{s=3}$ at the negative.
So we need to consider the correlator $<A_{s=3}(p_1)B_{s=4}(p_2)C(p_3)>$
where, for simplicity of notations, $C{\equiv}V_{s=3}$ is the $s=3$
unintegrated operator at picture $-3$ while 
$A_{s=3}$ and $B_{s=4}$ are given by (46) and (51). Using the decompositions
(47) and (51) - (54), simple analysis of ghost number balance shows
  that    $<A_{s=3}B_{s=4}C>$ is contributed by the correlators
\eqn\grav{\eqalign{<A_{s=3}B_{s=4}C>=
<A_0B_7C>+<A_6B_0C>+\sum_{i=1}^5\sum_{j=1}^6<A_iB_jC>}}
while all other correlators (e.g. such as $<A_0{B_j}C>,j=1,...,6$)
vanish due to the total ghost number constraints.
Below we shall perform the computation of the correlators (62), one by one.
We start with $<A_0{B_7}C>$. Using the expressions (1), (46) - (54)
for the operators and performing conformal mapping of the worldsheet
to the upper half plane (so the operators are located at the worldsheet
boundary which is the real axis) this correlation function
is given by
\eqn\grav{\eqalign{<A_0(p_1;z_1){B_7}(p_2;w_1)C(p_3,u)>
\cr
=-{1\over2}\times{12}H_{a_1a_2a_3}(p_1)H_{b_1...b_4}(p_2)H_{c_1c_2c_3}(p_3)
\int_0^1{dw}{\int{dz}}_{0\leq{z}<{w}}(z-z_1)^2(w-w_1)^2\cr\lbrace
<P^{(2)}_{2\phi-2\chi-\sigma}
e^{\phi}\partial{X^{a_1}}\partial{X^{a_2}}
\psi^{a_3}{e^{i{\vec{p_1}}{\vec{X}}}}(z)
\partial{c}c\xi\partial{X^{b_1}}
\partial{X^{b_2}}\partial\psi^{b_3}\psi^{b_4}{e^{i{\vec{p_2}}{\vec{X}}}}
(w)
\cr
ce^{-3\phi}\partial{X^{c_1}}\partial{X^{c_2}}\psi^{c_3}
{e^{i{\vec{p_2}}{\vec{X}}}}(u)>\rbrace}}
Using the $SL(2,R)$ symmetry, it is convenient
to set $z_1=0,w_1=1,u\rightarrow\infty$.
Note that $SL(2,R)$ symmetry is equivalent to the fact that
$z_1,w_1$ and $u$ derivatives of the vertex operators (36), (46)-(54)
are BRST-trivial (so that the correlation function
is invariant under the change of the operator's locations).
On the other hand, due to the ghost structure
of the higher spin vertices, the 3-point function
already contains 2 out of 3 integrated operators
despite gauge fixing $SL(2,R)$.
This is in contrast with the
standard case when the $SL(2,R)$ symmetry ensures that
all the operators in 3-point function s are unintegrated,
leading to the usual Koba-Nielsen's determinant.
The correlator in the integrand of (63) is the direct product of the
$\psi$, $X$ and ghost correlators.
Using the symmetries in $b_3$ and $b_4$ indices (since all the H-tensors,
including the one of $s=4$, are fully symmetric)
the $\psi$ part can be written as

\eqn\grav{\eqalign{<\psi^{a_3}(z)\partial\psi^{b_3}\psi^{b_4}(w)\psi^{c_3}(u)
>{\equiv}{1\over2}
<\psi^{a_3}(z)\partial\psi^{(b_3}\psi^{b_4)}(w)\psi^{c_3}(u)
>\cr
={1\over2}\eta^{a_3(b_3}\eta^{b_4)c_3}({{1}\over{(z-w)^2(w-u)}}
+{1\over{(z-w)(w-u)^2}})={1\over2}{{\eta^{a_3(b_3}\eta^{b_4)c_3}(z-u)}\over
{(z-w)^2(w-u)^2}}}} 
In the limit $u\rightarrow\infty$ we have
\eqn\lowen{
lim_{u\rightarrow{\infty}}{1\over2}<\psi^{a_3}(z)\partial\psi^{(b_3}
\psi^{b_4)}(w)\psi^{c_3}(u)>=
-{1\over2}{{\eta^{a_3(b_3}\eta^{b_4)c_3}}\over
{(z-w)^2{u}}}}
Next, the ghost part of the correlator (63) is given by
\eqn\grav{\eqalign{
<P^{(2)}_{2\phi-2\chi-\sigma}e^\phi(z)\partial{c}{c}\xi(w)c{e^{-3\phi}}(u)>
\cr
=(z-u)^3(w-u)^2({{4}\over{(z-w)^2}}-{{5}\over{(z-u)^2}}-{{40}\over{(z-w)(z-u)}})
}}
where we used 
\eqn\grav{\eqalign{P^{(2)}_{2\phi-2\chi-\sigma}={\partial}
P^{(1)}_{2\phi-2\chi-\sigma}+(P^{(1)}_{2\phi-2\chi-\sigma})^2}}
along with the OPEs:
\eqn\grav{\eqalign{P^{(1)}_{2\phi-2\chi-\sigma}(z)\partial{c}{c}\xi(w)
\sim{-}{{4\partial{c}{c}\xi}\over{z-w}}\cr
P^{(1)}_{2\phi-2\chi-\sigma}(z)ce^{-3\phi}(w)
\sim{{5ce^{-3\phi}}\over{z-w}}}}
In the limit $u\rightarrow\infty$ the ghost correlator becomes
\eqn\grav{\eqalign{lim_{u\rightarrow\infty}
<P^{(2)}_{2\phi-2\chi-\sigma}e^\phi(z)\partial{c}{c}\xi(w)c{e^{-3\phi}}(u)>
=-u^5({{4}\over{(z-w)^2}}-{5\over{u^2}}+{{40}\over{(z-w)u}})}}
Therefore in the limit $u\rightarrow\infty$
the ghost$\times\psi$-part of the correlator (63)
is expressed as
\eqn\grav{\eqalign{
<\psi^{a_3}(z)\partial\psi^{b_3}\psi^{b_4}(w)\psi^{c_3}(u)
><P^{(2)}_{2\phi-2\chi-\sigma}e^\phi(z)\partial{c}{c}\xi(w)c{e^{-3\phi}}(u)>
\cr
={u^4{\eta^{a_3(b_3}\eta^{b_4)c_3}}\over
{2(z-w)^2}}({{4}\over{(z-w)^2}}-{5\over{u^2}}+{{40}\over{(z-w)u}})}}
The only remaining part to compute is the
 $X$-correlator given by
$${<\partial{X^{a_1}}\partial{X^{a_2}}
{e^{i{\vec{p_1}}{\vec{X}}}}(z)\partial{X^{b_1}}
\partial{X^{b_2}}{e^{i{\vec{p_2}}{\vec{X}}}}
(w)\partial{X^{c_1}}\partial{X^{c_2}}
{e^{i{\vec{p_3}}{\vec{X}}}}(z)>}$$
This correlator has the structure (56) and is thus given by
the combination of 
$$T_{p,q,r|s}^{a_1...a_p|b_1...b_q|c_1...c_r}(p_1,p_2,p_3)$$
with $p=q=r=2$ but with the different values of $s$. There is, however,
a considerable simplification due to the conformal invariance of the theory.
Namely, consider the $\psi$-ghost part (70) of the correlator (63)
which, in the limit  $u\rightarrow\infty$ ,is given by
the order 4 polynomial containing positive powers of $u$.
On the other hand, the conformal invariance
only allows the terms behaving as $u^0$ when $u\rightarrow\infty$;
terms with the asymptotics ${\sim}u^n,n>0$ cannot appear 
on-shell,
as they are prohibited
by the conformal invariance; terms of the type ${1\over{u^n}},n>0$
vanish as $u\rightarrow\infty$. This condition very much limits
the on-shell contributions from the $X$-correlator received by
the non-vanishing terms in the overall correlator.
That is, note that, given the $\psi$-ghost factor (70)
the overall correlator has the following
structure at $u\rightarrow\infty$:
\eqn\grav{\eqalign{<A_0(p_1;z_1){B_7}(p_2;w_1)C(p_3,u)>
\cr
\sim
{{\eta^{a_3(b_3}\eta^{b_4)c_3}}\over
{2}}({{4u^4}\over{(z-w)^4}}-{{{5u^2}\over{(z-w)^2}}}+
{{40u^3}\over{(z-w)^3}})\sum_s
T_{p,q,r|s}^{a_1...a_p|b_1...b_q|c_1...c_r}(p_1,p_2,p_3)}}
Since by definition
$lim_{u\rightarrow\infty}T_{p,q,r|s}^{a_1...a_p|b_1...b_q|c_1...c_r}(p_1,p_2,p_3)
\sim{u^{-s}}$,
it is clear that the only non-vanishing contribution
picked from the $X$-correlator
is the one with $s=4$, i.e. with the value of $s$ equal to the
leading order of the $u$-asymptotics of the $\psi$-ghost factor
(here and in a number of places below, $s$ refers to the order
of the asymptotics and  not the spin value, we hope that
the difference shall be clear to the reader from the context). 
Those with $s<{4}$ are prohibited
by the conformal invariance (since they lead to
positive powers of $u$ in the asymptotics) and thus we know
in advance that they
must vanish on-shell;  those with $s>4$ are gauged
away in the limit $u\rightarrow\infty$.
Therefore substituting (64), (66), (71) in the integral (63) we obtain the  
following expression for the overall correlator:
\eqn\grav{\eqalign{<A_0(p_1){B_7}(p_2)C(p_3)>
\cr
=-24I({\vec{p}}_1{\vec{p}}_2)
{{\eta^{a_3b_3}\eta^{b_4c_3}}}T_{2,2,2|4}^{a_1a_2|b_1b_2|c_1c_2}
(p_1,p_2,p_3)
\cr\times
H_{a_1a_2a_3}(p_1)H_{b_1...b_4}(p_2)H_{c_1c_2c_3}(p_3)
\delta({\vec{p}}_1+{\vec{p}}_2+{\vec{p}}_3)}}

where

\eqn\grav{\eqalign{I({\vec{p}}_1{\vec{p}}_2)=
{{\int{dw}\int{dz}}_{0\leq{z}<w\leq{1}}}
{z^2}(w-1)^2(z-w)^{({\vec{p}}_1{\vec{p}}_2)-4}}}

The integral (73) is easy to evaluate.
We have:
\eqn\grav{\eqalign{I({\vec{p}}_1{\vec{p}}_2)
=\int_0^1{dw}(w-1)^2\int_0^w{dz}{z^2}(z-w)^{({\vec{p}}_1{\vec{p}}_2)-4}
\cr
=\int_{0}^1{dw}(w-1)^2{w^{({\vec{p}}_1{\vec{p}}_2)-4}}
\int_0^w{dz}z^2({z\over{w}}-1)^{({\vec{p}}_1{\vec{p}}_2)-4}\cr
=\int_{0}^1{dw}(w-1)^2{w^{({\vec{p}}_1{\vec{p}}_2)-1}}
\int_0^1{dx}{x^2}(x-1)^{({\vec{p}}_1{\vec{p}}_2)-4}
\cr
={{\Gamma(3)\Gamma({({\vec{p}}_1{\vec{p}}_2)-3})}\over
{\Gamma({({\vec{p}}_1{\vec{p}}_2)})}}\times
{{\Gamma(3)\Gamma({({\vec{p}}_1{\vec{p}}_2)})}\over
{\Gamma({({\vec{p}}_1{\vec{p}}_2)+3})}}
=4\prod_{n=-3}^2{1\over{{({\vec{p}}_1{\vec{p}}_2)+n}}}}}
where in the process we changed the integration variable
$x={{z\over{w}}}$.
In the on-shell limit one has $({\vec{p}}_1{\vec{p}}_2)\rightarrow{0}$
and the integral becomes 
\eqn\lowen{I({\vec{p}}_1{\vec{p}}_2)\approx
-{1\over{3({\vec{p}}_1{\vec{p}}_2)}}-{1\over9}+O({\vec{p}}_1{\vec{p}}_2)}
with the first term reflecting the non-localites well-known in the theories
of higher spins (e.g. ~{\sorokin, \deser, \mva,\mvd
\sagnottia,\sagnottif})
 and the second term corresponding to the local
part of the cubic interactions.
Note that the nonlocalities appear as a result of the ghost
structure of the vertex operators (36), (46) - (54) leading to appearance of the
integrated vertices in the three-point function and thus the deformation
of the standard Koba-Nielsen's measure.
The interaction terms in the position space are straightforward
to obtain by the Fourier transform. For example,
using (72), (74) and the expression (61) 
for $T_{p,q,r|s}^{a_1...a_p|b_1...b_q|c_1...c_r}(p_1,p_2,p_3)$
 the cubic interaction term in the higher spin Lagrangian due 
to the correlator (63) is given by
\eqn\grav{\eqalign{\sim{-24}{{\eta^{a_3b_3}\eta^{b_4c_3}}}
I({\vec{\partial}}_1{\vec{\partial}}_2)
\sum_{{\lbrace{m}\rbrace},{\lbrace{n}\rbrace}}
{{(-1)^{n_1+n_2+n_3+m_1+m_2+m_4}}\over{\prod_{i=1}^3{n_i}!\prod_{j=1}^6{m_j}!}}
\prod_{k=1}^{n_1}\eta^{a_kc_k}\prod_{k=1}^{n_3}\eta^{a_{n_1+k}b_k}
\cr\times
\prod_{k=1}^{n_2}\eta^{b_{n_3+k}c_{n_2+k}}
\prod_{k=1}^{m_1}\prod_{k=1}^{m_5}\partial^{c_{n_1+n_2+k}}
\partial^{b_{n_2+n_3+l}}H_{a_1a_2a_3}
\prod_{k=1}^{m_2}\prod_{l=1}^{m_3}
\partial^{c_{n_1+n_2+m_1+k}}\partial^{a_{n_3+n_1+l}}
H_{b_1...b_4}
\cr
\prod_{k=1}^{m_4}\prod_{l=1}^{m_6}
\partial^{a_{n_3+n_1+m_3+k}}\partial^{b_{n_2+n_3+m_5+l}}
H_{c_1c_2c_3}
+Symm\lbrace(a_1,...,a_p);(b_1,...,b_q);(c_1,...,c_r)\rbrace}}
where, according the notation of (76), the space-time derivatives
$\partial_1$ and $\partial_2$ of $I({\vec{\partial}}_1{\vec{\partial}}_2)$
act on $H_{a_1a_2a_3}$ and $H_{b_1...b_4}$ respectively.
The cubic interaction terms corresponding to other correlators
in (62), that we shall consider below, can be obtained 
in a totally similar way.
We now turn to the next correlator contributing to the
cubic interaction, $<A_6B_0C>$ of (62).
The calculation using the vertex operators (1), (46)-(54)
is totally similar to the one described above.
The result is given by
\eqn\grav{\eqalign{<A_6(p_1)B_0(p_2)C(p_3)>
\cr=
-24I({\vec{p}}_1{\vec{p}}_2)
{{\eta^{a_3b_3}\eta^{b_4c_3}}}T_{2,2,2|4}^{a_1a_2|b_1b_2|c_1c_2}
(p_1,p_2)H_{a_1a_2a_3}(p_1)H_{b_1...b_4}(p_2)H_{c_1c_2c_3}(p_3)
\cr\times
\delta({\vec{p}}_1+{\vec{p}}_2+{\vec{p}}_3)}}

so the sum of the first two contributions to the cubic interaction
vertex is
\eqn\grav{\eqalign{<A_0(p_1)B_7(p_2)C(p_3)>+<A_6(p_1)B_0(p_2)C(p_3)>
\cr=
-48I({\vec{p}}_1{\vec{p}}_2)
{{\eta^{a_3b_3}\eta^{b_4c_3}}}T_{2,2,2|4}^{a_1a_2|b_1b_2|c_1c_2}
(p_1,p_2)H_{a_1a_2a_3}(p_1)H_{b_1...b_4}(p_2)H_{c_1c_2c_3}(p_3)
\cr\times\delta({\vec{p}}_1+{\vec{p}}_2+{\vec{p}}_3)}}
Finally, we need to analyze the correlators
of the type $<A_jB_jC>$ in (62) ($i=1,...,5;j=1,...,6$)
which have different ghost structure and a bit more cumbersome
structure of the matter part.
The calculation of these correlators is also performed
according to the same procedure as above, in the gauge
 $z_1=0,w_1=1,u=\infty$. Below we shall present 
the results for these correlators, one by one.
The correlator $<A_1B_1C>$ is given by
\eqn\grav{\eqalign{
<A_1(p_1)B_1(p_2)C(p_3)>=
4H_{a_1a_2a_3}(p_1)H_{b_1...b_4}(p_2)H_{c_1c_2c_3}(p_3)
\cr\times
\int_0^1{dw}\int_{0\leq{z}<w}{dz}{z^2}(w-1)^2\lbrace
<c\xi(z)c{e^\phi}(w)c{e^{-3\phi}}(u)>
\cr\times
<({\vec{\psi}}\partial{\vec{X}})\partial{X^{a_1}}\partial{X^{a_2}}
\psi^{a_3}{e^{i{\vec{p}}_1{\vec{X}}}}(z)
({\vec{\psi}}\partial{\vec{X}})\partial{X^{b_1}}\partial{X^{b_2}}
\partial\psi^{b_3}\psi^{b_4}{e^{i{\vec{p}}_2{\vec{X}}}}(w)
\cr\times
\partial{X^{c_1}}\partial{X^{c_2}}
\psi^{c_3}{e^{i{\vec{p}}_3{\vec{X}}}}(u)>\rbrace\cr\equiv
4H_{a_1a_2a_4}(p_1)H_{b_1b_2b_4b_5}(p_2)H_{c_1c_2c_3}(p_3)
\cr\times
<c\xi(z)c{e^\phi}(w)c{e^{-3\phi}}(u)>
<\psi^{a_3}\psi^{a_4}(z)\psi^{b_3}\partial\psi^{b_4}\psi^{b_5}(w)
\psi^{c_3}(u)>
\cr\times
<\partial{X^{a_1}}\partial{X^{a_2}}\partial{X^{a_3}}
{e^{i{\vec{p}}_1{\vec{X}}}}(z)\partial{X^{b_1}}
\partial{X^{b_2}}\partial{X^{b_3}}
{e^{i{\vec{p}}_2{\vec{X}}}}(w)
\partial{X^{c_1}}\partial{X^{c_2}}
{e^{i{\vec{p}}_3{\vec{X}}}}(u)>\cr
=4I({\vec{p}}_1{\vec{p}}_2)\eta^{b_3\lbrack{a_3}}\eta^{a_4\rbrack{b_4}}
\eta^{c_3b_5}T_{3,3,2|4}^{a_1a_2a_3|b_1b_2b_3|c_1c_2}(p_1,p_2,p_3)
\cr\times
H_{a_1a_2a_4}(p_1)H_{b_1b_2b_4b_5}(p_2)H_{c_1c_2c_3}(p_3)}}
where, in order to mantain the order of indices
in $T_{3,3,2|4}^{a_1a_2a_3|b_1b_2b_3|c_1c_2}$, consistent with the expression
(61), we have renamed the indices in some of the contractions
(e.g. $b_3\rightarrow{b_4}$ $b_4\rightarrow{b_5}$ while reserving
$b_3$ for the contraction of $\psi$ and $\partial{X}$ in the second
vertex operator).
In addition, here and elsewhere below we shall suppress
the common factor of $\delta({\vec{p}}_1+{\vec{p}}_2+{\vec{p}}_3)$
in all the amplitudes. 
The next correlator to consider is $<A_1(p_1)(B_2+B_3)(p_2)C(p_3)>$.
Using the expressions (1), (46) - (54) for the corresponding operators, we have:
\eqn\grav{\eqalign{
<A_1(p_1)(B_2+B_3)(p_2)C(p_3)>=-4ip_2^{b_5}H_{a_1a_2a_4}(p_1)
H_{b_1b_2b_3b_4}(p_2)H_{c_1c_2c_3}(p_3)
\cr\times
\int_0^1{dw}\int_{0\leq{z}<w}{dz}{z^2}(w-1)^2\lbrace
<c\xi\psi^{a_3}\psi^{a_4}(z)c{e^\phi}\partial\psi^{b_3}\psi^{b_4}
(\psi^{b_5}P^{(1)}_{\phi-\chi}+\partial\psi^{b_5})(w)\psi^{c_3}(u)>
\cr\times
<\partial{X^{a_1}}\partial{X^{a_2}}\partial{X^{a_3}}
{e^{i{\vec{p}}_1{\vec{X}}}}(z)\partial{X^{b_1}}
\partial{X^{b_2}}
{e^{i{\vec{p}}_2{\vec{X}}}}(w)
\partial{X^{c_1}}\partial{X^{c_2}}
{e^{i{\vec{p}}_3{\vec{X}}}}(u)>\rbrace
\cr=
8i{I({\vec{p}}_1{\vec{p}}_2)}
p_2^{\lbrack{a_3}}\eta^{a_4\rbrack{b_4}}\eta^{b_3c_3}
T_{3,2,2|4}^{a_1a_2a_3|b_1b_2|c_1c_2}(p_1,p_2,p_3)
H_{a_1a_2a_4}(p_1)
H_{b_1b_2b_3b_4}(p_2)H_{c_1c_2c_3}(p_3)}}
The next correlator, $<A_1(p_1)B_4(p_2)C(p_3)>$, vanishes:
\eqn\grav{\eqalign{<A_1(p_1)B_5(p_2)C(p_3)>
\cr
=-4H_{a_1a_2a_4}(p_1)
H_{b_1b_2b_3b_4}(p_2)H_{c_1c_2c_3}(p_3)
\int_0^1{dw}\int_{0\leq{z}<w}{dz}{z^2}(w-1)^2
\cr\times\lbrace
<c\xi(z)P^{(2)}_{\phi-\chi}ce^\phi(w)ce^{-3\phi}(u)>
<\psi^{a_3}\psi^{a_4}(z)\partial^2\psi^{b_2}\partial\psi^{b_3}\psi^{b_4}(w)
\psi^{c_3}(u)>
\cr\times
<\partial{X^{a_1}}\partial{X^{a_2}}\partial{X^{a_3}}
{e^{i{\vec{p}}_1{\vec{X}}}}(z)\partial{X^{b_1}}
{e^{i{\vec{p}}_2{\vec{X}}}}(w)
\partial{X^{c_1}}\partial{X^{c_2}}
{e^{i{\vec{p}}_3{\vec{X}}}}(u)>\rbrace
 =0}}
due to the vanishing $\psi$-correlator
(which is equal to zero as  the fermions
$\psi^{a_3}\psi^{a_4}$ at $z$, antisymmetric in
the $a_3,a_4$ indices
always contract with 2 out of 3 fermions
$\partial^2\psi^{b_2}\partial\psi^{b_3}\psi^{b_4}$  at $w$
which are  symmetric in $b_2,b_3,b_4$ (since
$H_{b_1...b_4}$ is totally symmetric)
The next correlator,
\eqn\grav{\eqalign{
<A_1(p_1)B_5(p_2)C(p_3)>=
-4H_{a_2a_3a_4}(p_1)
H_{b_1b_2b_3b_4}(p_2)H_{c_1c_2c_3}(p_3)
\cr\times
\int_0^1{dw}\int_{0\leq{z}<w}{dz}{z^2}(w-1)^2
\lbrace
<\psi^{a_1}(z)\psi^{a_4}\psi^{b_4}(w)\psi^{c_3}(u)>
\cr\times
<c\xi
\partial{X^{a_1}}\partial{X^{a_2}}\partial{X^{a_3}}
{e^{i{\vec{p}}_1{\vec{X}}}}(z)
ce^\phi\partial{X^{b_1}}\partial{X^{b_2}}
({1\over2}\partial^3{X^{b_3}}+
\partial^2{X^{b_3}}P^{(1)}_{\phi-\chi}
\cr
+{1\over2}\partial{X^{b_3}}P^{(2)}_{\phi-\chi})
{e^{i{\vec{p}}_2{\vec{X}}}}(w)
\partial{X^{c_1}}\partial{X^{c_2}}
{e^{i{\vec{p}}_3{\vec{X}}}}(u)>\rbrace
\cr=
{I({\vec{p}}_1{\vec{p}}_2)}\eta^{b_4\lbrack{a_4}}\eta^{a_1\rbrack{c_3}}
\lbrace{48}\eta^{a_3b_3}T_{2,2,2|4}^{a_1a_2|b_1b_2|c_1c_2}(p_1,p_2,p_3)
+8\eta^{b_3c_2}T_{3,2,1|2}^{a_1a_2a_3|b_1b_2|c_1}(p_1,p_2,p_3)
\cr
-16ip_1^{b_3}T_{3,2,2|4}^{a_1a_2a_3|b_1b_2|c_1c_2}(p_1,p_2,p_3)
-4ip_3^{b_3}T_{3,2,2|3}^{a_1a_2a_3|b_1b_2|c_1c_2}(p_1,p_2,p_3)\rbrace
\cr\times
H_{a_2a_3a_4}(p_1)H_{b_1b_2b_3b_4}(p_2)H_{c_1c_2c_3}(p_3)}}
The next correlator,
\eqn\grav{\eqalign{
<A_1(p_1)B_6(p_2)C(p_3)>=
4H_{a_2a_3a_4}(p_1)
H_{b_1b_2b_3b_4}(p_2)H_{c_1c_2c_3}(p_3)
\cr\times
\int_0^1{dw}\int_{0\leq{z}<w}{dz}{z^2}(w-1)^2\lbrace
<\psi^{a_1}(z)\psi^{a_4}\partial\psi^{b_4}(w)\psi^{c_3}(u)>
\cr\times
<c\xi
\partial{X^{a_1}}\partial{X^{a_2}}\partial{X^{a_3}}
{e^{i{\vec{p}}_1{\vec{X}}}}(z)
ce^\phi\partial{X^{b_1}}\partial{X^{b_2}}
(\partial^2{X^{b_3}}
\cr
+\partial{X^{b_3}}P^{(1)}_{\phi-\chi})
{e^{i{\vec{p}}_2{\vec{X}}}}(w)
ce^{-3\phi}\partial{X^{c_1}}\partial{X^{c_2}}
{e^{i{\vec{p}}_3{\vec{X}}}}(u)>\rbrace
\cr
={I({\vec{p}}_1{\vec{p}}_2)}\eta^{b_4\lbrack{a_4}}\eta^{a_1\rbrack{c_3}}
\lbrace{-24}\eta^{a_3b_3}T_{2,2,2|4}^{a_1a_2|b_1b_2|c_1c_2}(p_1,p_2,p_3)
\cr
+8\eta^{b_3c_2}T_{3,2,1|2}^{a_1a_2a_3|b_1b_2|c_1}(p_1,p_2,p_3)
+8ip_1^{b_3}T_{3,2,2|4}^{a_1a_2a_3|b_1b_2|c_1c_2}(p_1,p_2,p_3)
\cr
+4ip_3^{b_3}T_{3,2,2|3}^{a_1a_2a_3|b_1b_2|c_1c_2}(p_1,p_2,p_3)\rbrace
H_{a_2a_3a_4}(p_1)H_{b_1b_2b_3b_4}(p_2)H_{c_1c_2c_3}(p_3)
}}
and therefore
\eqn\grav{\eqalign{
<A_1(p_1)(B_5+B_6)(p_2)C(p_3)>=\cr
{I({\vec{p}}_1{\vec{p}}_2)}\eta^{b_4\lbrack{a_4}}\eta^{a_1\rbrack{c_3}}
\lbrace{24}\eta^{a_3b_3}T_{2,2,2|4}^{a_1a_2|b_1b_2|c_1c_2}(p_1,p_2,p_3)
\cr
+16\eta^{b_3c_2}T_{3,2,1|2}^{a_1a_2a_3|b_1b_2|c_1}(p_1,p_2,p_3)
-8ip_1^{b_3}T_{3,2,2|4}^{a_1a_2a_3|b_1b_2|c_1c_2}(p_1,p_2,p_3)
\rbrace\cr\times
H_{a_2a_3a_4}(p_1)H_{b_1b_2b_3b_4}(p_2)H_{c_1c_2c_3}(p_3)
}}
This concludes the list of all the correlators involving $A_1(p_1)$.
Next,
\eqn\grav{\eqalign{
<(A_2+A_3)(p_1)B_1(p_2)C(p_3)>=
-4ip_1^{a_4}H_{a_1a_2a_3}(p_1)
H_{b_1b_2b_4b_5}(p_2)H_{c_1c_2c_3}(p_3)
\cr\times
\int_0^1{dw}\int_{0\leq{z}<w}{dz}z^2(w-1)^2\lbrace
<c\xi(\partial\psi^{a_4}+\psi^{a_4}P^{(1)}_{\phi-\chi})\psi^{a_3}
\partial{X^{a_1}}\partial{X^{a_2}}
{e^{i{\vec{p}}_1{\vec{X}}}}(z)
\cr
c{e^\phi}\psi^{b_3}\partial\psi^{b_4}\psi^{b_5}
\partial{X^{b_1}}\partial{X^{b_2}}
{e^{i{\vec{p}}_2{\vec{X}}}}(w)
\cr
ce^{-3\phi}\partial{X^{c_1}}\partial{X^{c_2}}\psi^{c_3}
{e^{i{\vec{p}}_3{\vec{X}}}}(u)>\rbrace
=
4i{I({\vec{p}}_1{\vec{p}}_2)}p_1^{a_4}
(-5\eta^{b_3c_3}\eta^{a_3b_4}\eta^{a_4b_5}+
3\eta^{a_3b_3}\eta^{b_4c_3}\eta^{a_4b_5}
\cr
+2\eta^{a_4b_3}\eta^{a_3b_5}\eta^{b_4c_3})
T_{2,3,2|4}^{a_1a_2|b_1b_2b_3|c_1c_2}(p_1,p_2,p_3)
H_{a_1a_2a_3}(p_1)
H_{b_1b_2b_4b_5}(p_2)H_{c_1c_2c_3}(p_3)}}
The next correlator is
\eqn\grav{\eqalign{
<(A_2+A_3)(p_1)(B_2+B_3)(p_2)C(p_3)>
\cr=
-4p_1^{a_4}p_2^{b_5}
H_{a_1a_2a_3}(p_1)
H_{b_1b_2b_3b_4}(p_2)H_{c_1c_2c_3}(p_3)
\int_0^1{dw}\int_{0\leq{z}<w}{dz}{z^2}(w-1)^2\cr\lbrace
<c\xi(\partial\psi^{a_4}+\psi^{a_4}P^{(1)}_{\phi-\chi})\psi^{a_3}(z)
ce^\phi\partial\psi^{b_3}\psi^{b_4}
(\psi_{b_5}P^{(1)}_{\phi-\chi}+\partial\psi^{b_5})(w)
ce^{-3\phi}
\psi^{c_3}(u)>
\cr{\times}
<\partial{X^{a_1}}\partial{X^{a_2}}
{e^{i{\vec{p}}_1{\vec{X}}}}(z)
\partial{X^{b_1}}\partial{X^{b_2}}
{e^{i{\vec{p}}_2{\vec{X}}}}(w)\partial{X^{c_1}}
\partial{X^{c_2}}\psi^{c_3}
{e^{i{\vec{p}}_3{\vec{X}}}}(u)>\rbrace
\cr=
-4{I({\vec{p}}_1{\vec{p}}_2)}
p_1^{b_4}p_2^{b_5}(5\eta^{a_3b_4}\eta^{a_4b_5}\eta^{b_3c_3}
-6\eta^{a_4b_3}\eta^{a_3b_5}\eta^{b_4c_3}
+\eta^{a_3b_4}\eta^{a_4b_3}\eta^{b_5c_3})
\cr\times
T_{2,2,2|4}^{a_1a_2|b_1b_2|c_1c_2}(p_1,p_2,p_3)
H_{a_1a_2a_3}(p_1)
H_{b_1b_2b_3b_4}(p_2)H_{c_1c_2c_3}(p_3)
}}

Next,
\eqn\grav{\eqalign{
<(A_2+A_3)(p_1)B_4(p_2)C(p_3)>
\cr
=4ip_1^{a_4}
H_{a_1a_2a_3}(p_1)
H_{b_1b_2b_3b_4}(p_2)H_{c_1c_2c_3}(p_3)
\int_0^1{dw}\int_{0\leq{z}<w}{dz}{z^2}(w-1)^2\cr\lbrace
<c\xi(\partial\psi^{a_4}+\psi^{a_4}P^{(1)}_{\phi-\chi})\psi^{a_3}(z)
ce^\phi\partial^2\psi^{b_2}\partial\psi^{b_3}\psi^{b_4}(w)
ce^{-3\phi}\psi^{c_3}(u)>
\cr{\times}
<\partial{X^{a_1}}\partial{X^{a_2}}
{e^{i{\vec{p}}_1{\vec{X}}}}(z)
\partial{X^{b_1}}
{e^{i{\vec{p}}_2{\vec{X}}}}(w)\partial{X^{c_1}}
\partial{X^{c_2}}
{e^{i{\vec{p}}_3{\vec{X}}}}(u)>\rbrace
\cr=
-16i{I({\vec{p}}_1{\vec{p}}_2)}
p_1^{a_4}\eta^{a_3b_3}\eta^{a_4b_2}\eta^{b_4c_3}
T_{2,1,2|4}^{a_1a_2|b_1|c_1c_2}(p_1,p_2,p_3)
\cr\times
H_{a_1a_2a_3}(p_1)
H_{b_1b_2b_3b_4}(p_2)H_{c_1c_2c_3}(p_3)}}
Next,
\eqn\grav{\eqalign{
<(A_2+A_3)(p_1)B_5(p_2)C(p_3)>
=4ip_1^{a_4}
H_{a_1a_2a_3}(p_1)
H_{b_1b_2b_3b_4}(p_2)H_{c_1c_2c_3}(p_3)
\cr\times
\int_0^1{dw}\int_{0\leq{z}<w}{dw}z^2(w-1)^2\lbrace
<c\xi(\partial\psi^{a_4}+\psi^{a_4}P^{(1)}_{\phi-\chi})\psi^{a_3}(z)
\partial{X^{a_1}}\partial{X^{a_2}}
{e^{i{\vec{p}}_1{\vec{X}}}}(z)
\cr
ce^{\phi}\psi^{b_4}\partial{X^{b_1}}\partial{X^{b_2}}
({1\over2}\partial^3{X^{b_3}}+
\partial^2{X^{b_3}}P^{(1)}_{\phi-\chi}
+{1\over2}\partial{X^{b_3}}P^{(2)}_{\phi-\chi})
{e^{i{\vec{p}}_2{\vec{X}}}}(w)
\cr
ce^{-3\phi}\psi^{c_3}\partial{X^{c_1}}
\partial{X^{c_2}}
{e^{i{\vec{p}}_3{\vec{X}}}}(u)>\cr
={I({\vec{p}}_1{\vec{p}}_2)}\lbrace
{48i}(p_1^{c_3}\eta^{a_3b_4}\eta^{b_3a_2}
-2p_1^{b_4}\eta^{a_2b_3}\eta^{a_3c_3})
T_{1,2,2|4}^{a_1|b_1b_2|c_1c_2}(p_1,p_2,p_3)
\cr
+8i(p_1^{c_3}\eta^{a_3b_4}\eta^{b_3c_2}
-2p_1^{b_4}\eta^{a_3c_3}\eta^{b_3c_2})
T_{2,2,1|2}^{a_1a_2|b_1b_2|c_1}(p_1,p_2,p_3)
\cr
+4p_1^{c_3}p_3^{b_3}\eta^{a_3b_4}
T_{2,2,2|3}^{a_1a_2|b_1b_2|c_1c_2}(p_1,p_2,p_3)
-16p_1^{c_3}p_1^{b_3}\eta^{a_3b_4}
T_{2,2,2|4}^{a_1a_2|b_1b_2|c_1c_2}(p_1,p_2,p_3)
\cr
-32p_1^{b_3}p_1^{b_4}\eta^{a_3c_3}
T_{2,2,2|4}^{a_1a_2|b_1b_2|c_1c_2}(p_1,p_2,p_3)
-8p_3^{b_3}p_1^{b_4}\eta^{a_3c_3}
T_{2,2,2|3}^{a_1a_2|b_1b_2|c_1c_2}(p_1,p_2,p_3)
\cr\times
H_{a_1a_2a_3}(p_1)
H_{b_1b_2b_3b_4}(p_2)H_{c_1c_2c_3}(p_3)
}}

The last correlator involving $(A_2+A_3)(p_1)$ is
\eqn\grav{\eqalign{
<(A_2+A_3)(p_1)B_6(p_2)C(p_3)>
=-4ip_1^{a_4}
H_{a_1a_2a_3}(p_1)
H_{b_1b_2b_3b_4}(p_2)H_{c_1c_2c_3}(p_3)
\cr\times
\int_0^1{dw}\int_{0\leq{z}<w}{dw}z^2(w-1)^2\lbrace
<c\xi(\partial\psi^{a_4}+\psi^{a_4}P^{(1)}_{\phi-\chi})\psi^{a_3}(z)
\partial{X^{a_1}}\partial{X^{a_2}}
{e^{i{\vec{p}}_1{\vec{X}}}}(z)
\cr
ce^{\phi}\partial\psi^{b_4}\partial{X^{b_1}}\partial{X^{b_2}}
(\partial^2{X^{b_3}}+\partial{X^{b_3}}P^{(1)}_{\phi-\chi})
ce^{-3\phi}\psi^{c_3}\partial{X^{c_1}}
\partial{X^{c_2}}
{e^{i{\vec{p}}_3{\vec{X}}}}(u)>
\cr
={I({\vec{p}}_1{\vec{p}}_2)}\lbrace(
72ip_1^{b_4}\eta^{a_2b_3}\eta^{a_3c_3}-24ip_1^{c_3}\eta^{a_2b_3}\eta^{a_3b_4})
T_{1,2,2|4}^{a_1|b_1b_2|c_1c_2}(p_1,p_2,p_3)
\cr
+(24p_1^{b_3}p_1^{b_4}\eta^{a_3c_3}-8p_1^{b_3}p_1^{c_3}\eta^{a_3b_4})
T_{2,2,2|4}^{a_1a_2|b_1b_2|c_1c_2}(p_1,p_2,p_3)\cr
+(12p_1^{b_4}p_3^{b_3}\eta^{a_3c_3}-4p_1^{c_3}p_3^{b_3}\eta^{a_3b_4})
T_{2,2,2|3}^{a_1a_2|b_1b_2|c_1c_2}(p_1,p_2,p_3)
\cr
+(24ip_1^{b_4}\eta^{a_3c_3}\eta^{b_3c_2}-8ip_1^{c_3}
\eta^{a_3b_4}\eta^{b_3c_2})T_{2,2,1|2}^{a_1a_2|b_1b_2|c_1}(p_1,p_2,p_3)
\cr\times
H_{a_1a_2a_3}(p_1)
H_{b_1b_2b_3b_4}(p_2)H_{c_1c_2c_3}(p_3)
}}
Next, we shall consider the correlators with $A_4(p_1)$.
We have:
\eqn\grav{\eqalign{
<A_4(p_1)B_1(p_2)C(p_3)>=-4H_{a_1a_2a_3}(p_1)
H_{b_1b_2b_4b_5}(p_2)H_{c_1c_2c_3}(p_3)
\cr\times
\int_0^1{dw}\int_{0\leq{z}<w}{dw}z^2(w-1)^2\lbrace
<c\xi\partial{X^{a_1}}
(\partial^2\psi^{a_2}+2\partial\psi^{a_2}
P^{(1)}_{\phi-\chi})\psi^{a_3}{e^{i{\vec{p}}_1{\vec{X}}}}(z)
\cr
{ce^\phi}
({\vec{\psi}}\partial{\vec{X}})\partial{{X^{b_1}}}\partial{X^{b_2}}
\partial\psi^{b_4}\psi^{b_5}{e^{i{\vec{p}}_2{\vec{X}}}}(w)
ce^{-3\phi}\psi^{c_3}\partial{X^{c_1}}
\partial{X^{c_2}}
{e^{i{\vec{p}}_3{\vec{X}}}}(u)>\rbrace\cr
=24{I({\vec{p}}_1{\vec{p}}_2)}\eta^{a_2b_4}\eta^{a_3\lbrack{b_3}}
\eta^{{b_5}{\rbrack}c_3}T_{1,3,2|4}^{a_1|b_1b_2b_3|c_1c_2}(p_1,p_2,p_3)
H_{a_1a_2a_3}(p_1)
H_{b_1b_2b_4b_5}(p_2)H_{c_1c_2c_3}(p_3)}}
The next correlator is
\eqn\grav{\eqalign{
<A_4(p_1)(B_2+B_3)(p_2)C(p_3)>
=4ip_2^{b_5}H_{a_1a_2a_3}(p_1)
H_{b_1b_2b_3b_4}(p_2)H_{c_1c_2c_3}(p_3)
\cr\times
\int_0^1{dw}\int_{0\leq{z}<w}{dw}z^2(w-1)^2\lbrace
<c\xi\partial{X^{a_1}}
(\partial^2\psi^{a_2}+2\partial\psi^{a_2}
P^{(1)}_{\phi-\chi})\psi^{a_3}{e^{i{\vec{p}}_1{\vec{X}}}}(z)
\cr
{ce^\phi}\partial{X^{b_1}}\partial{X^{b_2}}\partial\psi^{b_3}
\psi^{b_4}(\partial\psi^{b_5}+\psi^{b_5}P^{(1)}_{\phi-\chi})
{e^{i{\vec{p}}_2{\vec{X}}}}(w)
ce^{-3\phi}\psi^{c_3}\partial{X^{c_1}}
\partial{X^{c_2}}
{e^{i{\vec{p}}_3{\vec{X}}}}(u)>\rbrace\cr
=-24ip_2^{b_5}
{I({\vec{p}}_1{\vec{p}}_2)}\lbrace
(\eta^{a_3b_3}\eta^{a_2b_4}\eta^{b_5c_3}
+\eta^{a_3b_3}\eta^{a_2b_5}\eta^{b_4c_3})
T_{1,2,2|4}^{a_1|b_1b_2|c_1c_2}(p_1,p_2,p_3)
\cr
+\eta^{a_3b_4}\eta^{a_2b_5}\eta^{b_3c_3}
T_{1,2,2|4}^{a_1|b_1b_2|c_1c_2}(p_1,p_2,p_3)
\rbrace
H_{a_1a_2a_3}(p_1)
H_{b_1b_2b_3b_4}(p_2)H_{c_1c_2c_3}(p_3)}}

Next,
\eqn\grav{\eqalign{
<A_4(p_1)B_4(p_2)C(p_3)>
=4H_{a_1a_2a_3}(p_1)
H_{b_1b_2b_3b_4}(p_2)H_{c_1c_2c_3}(p_3)
\cr\times
\int_0^1{dw}\int_{0\leq{z}<w}{dw}z^2(w-1)^2\lbrace
<c\xi\partial{X^{a_1}}
(\partial^2\psi^{a_2}+2\partial\psi^{a_2}
P^{(1)}_{\phi-\chi})\psi^{a_3}{e^{i{\vec{p}}_1{\vec{X}}}}(z)
\cr
{ce^\phi}P^{(2)}_{\phi-\chi}\partial{X^{b_1}}\partial^2\psi^{b_2}
\partial\psi^{b_3}\psi^{b_4}
ce^{-3\phi}\psi^{c_3}\partial{X^{c_1}}
\partial{X^{c_2}}
{e^{i{\vec{p}}_3{\vec{X}}}}(u)>\rbrace\cr=
128{I({\vec{p}}_1{\vec{p}}_2)}\eta^{a_2b_2}\eta^{a_3b_3}\eta^{b_4c_3}
T_{1,1,2|4}^{a_1|b_1|c_1c_2}(p_1,p_2,p_3)
H_{a_1a_2a_3}(p_1)
H_{b_1b_2b_3b_4}(p_2)H_{c_1c_2c_3}(p_3)}}
Next,
\eqn\grav{\eqalign{
<A_4(p_1)B_5(p_2)C(p_3)>
=4H_{a_1a_2a_3}(p_1)
H_{b_1b_2b_3b_4}(p_2)H_{c_1c_2c_3}(p_3)
\cr\times
\int_0^1{dw}\int_{0\leq{z}<w}{dw}z^2(w-1)^2\lbrace
<c\xi\partial{X^{a_1}}
(\partial^2\psi^{a_2}+2\partial\psi^{a_2}
P^{(1)}_{\phi-\chi})\psi^{a_3}{e^{i{\vec{p}}_1{\vec{X}}}}(z)
\cr
{ce^\phi}\partial{X^{b_1}}\partial{X^{b_2}}({1\over2}
\partial^3{X^{b_3}}+\partial^2{X^{b_3}}P^{(1)}_{\phi-\chi}
\cr
+{1\over2}\partial{X^{b_3}}P^{(2)}_{\phi-\chi})\psi^{b_4}
{e^{i{\vec{p}}_2{\vec{X}}}}(w)
ce^{-3\phi}\psi^{c_3}\partial{X^{c_1}}
\partial{X^{c_2}}
{e^{i{\vec{p}}_3{\vec{X}}}}(u)>\rbrace\cr=
16{I({\vec{p}}_1{\vec{p}}_2)}\eta^{a_2b_4}\eta^{a_3c_3}
\lbrace
6\eta^{a_1b_3}
T_{0,2,2|4}^{-|b_1b_2|c_1c_2}(p_1,p_2,p_3)
+2\eta^{b_3c_2}
T_{1,2,1|2}^{a_1|b_1b_2|c_1}(p_1,p_2,p_3)
\cr
+4ip_1^{b_3}T_{1,2,2|4}^{a_1|b_1b_2|c_1c_2}(p_1,p_2,p_3)
+ip_3^{b_3}T_{1,2,2|3}^{a_1|b_1b_2|c_1c_2}(p_1,p_2,p_3)\rbrace
\cr\times
H_{a_1a_2a_3}(p_1)
H_{b_1b_2b_3b_4}(p_2)H_{c_1c_2c_3}(p_3)}}
The last correlator involving $A_4(p_1)$ is
\eqn\grav{\eqalign{
<A_4(p_1)B_6(p_2)C(p_3)>
=-4H_{a_1a_2a_3}(p_1)
H_{b_1b_2b_3b_4}(p_2)H_{c_1c_2c_3}(p_3)
\cr\times
\int_0^1{dw}\int_{0\leq{z}<w}{dw}z^2(w-1)^2\lbrace
<c\xi\partial{X^{a_1}}
(\partial^2\psi^{a_2}+2\partial\psi^{a_2}
P^{(1)}_{\phi-\chi})\psi^{a_3}{e^{i{\vec{p}}_1{\vec{X}}}}(z)
\cr
{ce^\phi}\partial{X^{b_1}}\partial{X^{b_2}}
(\partial^2{X^{b_3}}+\partial{X^{b_3}}P^{(1)}_{\phi-\chi})
\partial\psi^{b_4}{e^{i{\vec{p}}_2{\vec{X}}}}(w)
ce^{-3\phi}\psi^{c_3}\partial{X^{c_1}}
\partial{X^{c_2}}
{e^{i{\vec{p}}_3{\vec{X}}}}(u)>\rbrace\cr=
16{I({\vec{p}}_1{\vec{p}}_2)}\eta^{a_2b_4}\eta^{a_3c_3}
\lbrace
-6\eta^{a_1b_3}T_{0,2,2|4}^{-|b_1b_2|c_1c_2}(p_1,p_2,p_3)
-4\eta^{b_3c_2}T_{1,2,1|2}^{a_1|b_1b_2|c_1}(p_1,p_2,p_3)
\cr
+2ip_1^{b_3}T_{1,2,2|4}^{a_1|b_1b_2|c_1c_2}(p_1,p_2,p_3)
+ip_3^{b_3}T_{1,2,2|3}^{a_1|b_1b_2|c_1c_2}(p_1,p_2,p_3)\rbrace
\cr\times
H_{a_1a_2a_3}(p_1)
H_{b_1b_2b_3b_4}(p_2)H_{c_1c_2c_3}(p_3)}}
and therefore
\eqn\grav{\eqalign{
<A_4(p_1)(B_5+B_6)(p_2)C(p_3)>
=
16{I({\vec{p}}_1{\vec{p}}_2)}\eta^{a_2b_4}\eta^{a_3c_3}\lbrace
-2\eta^{b_3c_2}T_{1,2,1|2}^{a_1|b_1b_2|c_1}(p_1,p_2,p_3)
\cr
+6ip_1^{b_3}T_{1,2,2|4}^{a_1|b_1b_2|c_1c_2}(p_1,p_2,p_3)
+2ip_3^{b_3}T_{1,2,2|3}^{a_1|b_1b_2|c_1c_2}(p_1,p_2,p_3)\rbrace
\cr\times
H_{a_1a_2a_3}(p_1)
H_{b_1b_2b_3b_4}(p_2)H_{c_1c_2c_3}(p_3)}}
This concludes the list of the correlators involving $A_4(p_1)$
Finally, we are left to consider the set of the correlators
involving $A_5(p_1)$.
Note that the expression for $A_5(p_1)$ contains no $\psi$'s at all
while the one for $C(p_3)$ has only one $\psi$.
At the same time, the operators
$B_1,B_2,B_3$ and $B_4$ are all cubic in $\psi$. For this reason,
\eqn\grav{\eqalign{<A_5(p_1)B_1(p_2)C(p_3)>=
<A_5(p_1)B_2(p_2)C(p_3)>
\cr
=<A_5(p_1)B_3(p_2)C(p_3)>
=<A_5(p_1)B_4(p_2)C(p_3)>=0}}
and the only non-vanishing correlators with $A_5$ are
$<A_5(p_1)B_5(p_2)C(p_3)>$ and $<A_5(p_1)B_6(p_2)C(p_3)>$.
For these remaining correlators we obtain
\eqn\grav{\eqalign{
<A_5(p_1)B_5(p_2)C(p_3)>
=-4H_{a_1a_2a_3}(p_1)
H_{b_1b_2b_3b_4}(p_2)H_{c_1c_2c_3}(p_3)
\cr\times
\int_0^1{dw}\int_{0\leq{z}<w}{dw}z^2(w-1)^2\lbrace
<c\xi\partial{X^{a_1}}\partial{X^{a_2}}(\partial^2{X^{a_3}}+
\partial{X^{a_3}}P^{(1)}_{\phi-\chi}){e^{i{\vec{p}}_1{\vec{X}}}}(z)
\cr
{ce^\phi}\partial{X^{b_1}}\partial{X^{b_2}}({1\over2}
\partial^3{X^{b_3}}+\partial^2{X^{b_3}}P^{(1)}_{\phi-\chi}
+{1\over2}\partial{X^{b_3}}P^{(2)}_{\phi-\chi})\psi^{b_4}
{e^{i{\vec{p}}_2{\vec{X}}}}(w)
\cr
ce^{-3\phi}\psi^{c_3}\partial{X^{c_1}}
\partial{X^{c_2}}
{e^{i{\vec{p}}_3{\vec{X}}}}(u)>\rbrace\cr=
4{I({\vec{p}}_1{\vec{p}}_2)}{\lbrace}
6\eta^{a_3b_2}\eta^{b_4c_3}(6\eta^{a_2b_3}
T_{1,1,2|4}^{a_1|b_1|c_1c_2}(p_1,p_2,p_3)
+6\eta^{b_3c_2}T_{2,1,1|2}^{a_1a_2|b_1|c_1}(p_1,p_2,p_3)
\cr
-4ip_1^{b_3}T_{2,1,2|4}^{a_1a_2|b_1|c_1c_2}(p_1,p_2,p_3)
+ip_3^{b_3}T_{2,1,2|3}^{a_1a_2|b_1|c_1c_2}(p_1,p_2,p_3))
\cr
+\eta^{a_3c_2}\eta^{b_4c_3}(-24\eta^{a_2b_3}
T_{1,2,1|2}^{a_1|b_1b_2|c_1}(p_1,p_2,p_3)
-8ip_1^{b_3}T_{2,2,1|2}^{a_1a_2|b_1b_2|c_1}(p_1,p_2,p_3)
\cr
-2ip_3^{b_3}T_{2,2,1|1}^{a_1a_2|b_1b_2|c_1}(p_1,p_2,p_3))
\cr
+26\eta^{a_3b_3}\eta^{b_4c_3}
T_{2,2,2|4}^{a_1a_2|b_1b_2|c_1c_2}(p_1,p_2,p_3)
+12\eta^{a_2b_3}\eta^{b_4c_3}(-2ip_2^{a_3}
T_{1,2,2|4}^{a_1|b_1b_2|c_1c_2}(p_1,p_2,p_3)
\cr
+ip_3^{a_3}T_{1,2,2|3}^{a_1|b_1b_2|c_1c_2}(p_1,p_2,p_3))
\cr
+2\eta^{b_4c_3}\eta^{b_3c_2}(-2ip_2^{a_3}
T_{2,2,1|2}^{a_1a_2|b_1b_2|c_1}(p_1,p_2,p_3)
+ip_3^{a_2}T_{2,2,1|1}^{a_1a_2|b_1b_2|c_1}(p_1,p_2,p_3))
\cr
+\eta^{b_4c_3}(-2p_2^{a_3}+p_3^{a_3})(4p_1^{b_3}
T_{2,2,2|4}^{a_1a_2|b_1b_2|c_1c_2}(p_1,p_2,p_3)
+p_3^{b_3}
T_{2,2,2|3}^{a_1a_2|b_1b_2|c_1c_2}(p_1,p_2,p_3))\rbrace
\cr\times
H_{a_1a_2a_3}(p_1)
H_{b_1b_2b_3b_4}(p_2)H_{c_1c_2c_3}(p_3)
}}
and finally
\eqn\grav{\eqalign{
<A_5(p_1)B_6(p_2)C(p_3)>
=4H_{a_1a_2a_3}(p_1)
H_{b_1b_2b_3b_4}(p_2)H_{c_1c_2c_3}(p_3)
\cr
\int_0^1{dw}\int_{0\leq{z}<w}{dw}z^2(w-1)^2\lbrace
<c\xi\partial{X^{a_1}}\partial{X^{a_2}}(\partial^2{X^{a_3}}+
\partial{X^{a_3}}P^{(1)}_{\phi-\chi}){e^{i{\vec{p}}_1{\vec{X}}}}(z)
\cr
ce^\phi\partial{X^{b_1}}\partial{X^{b_2}}
(\partial^2{X^{b_3}}
+\partial{X^{b_3}}P^{(1)}_{\phi-\chi})
\partial\psi^{b_4}
{e^{i{\vec{p}}_2{\vec{X}}}}(w)
ce^{-3\phi}\psi^{c_3}\partial{X^{c_1}}
\partial{X^{c_2}}
{e^{i{\vec{p}}_3{\vec{X}}}}(u)>\rbrace\cr=
4{I({\vec{p}}_1{\vec{p}}_2)}{\lbrace}-11\eta^{a_3b_3}\eta^{b_4c_3}
T_{2,2,2|3}^{a_1a_2|b_1b_2|c_1c_2}(p_1,p_2,p_3)
\cr
+\eta^{b_4c_3}\eta^{a_3b_2}(36\eta^{b_3a_2}
T_{1,1,2|3}^{a_1a_2|b_1b_2|c_1c_2}(p_1,p_2,p_3)
\cr
-12ip_1^{b_3}T_{2,1,2|3}^{a_1a_2|b_1|c_1c_2}(p_1,p_2,p_3)
-6ip_3^{b_3}T_{2,1,2|2}^{a_1a_2|b_1|c_1c_2}(p_1,p_2,p_3))
\cr
-12i\eta^{b_4c_3}\eta^{b_3a_2}p_2^{a_3}
T_{1,2,2|3}^{a_1|b_1b_2|c_1c_2}(p_1,p_2,p_3)
+4\eta^{b_4c_3}\eta^{b_3c_2}T_{2,2,1|1}^{a_1a_2|b_1b_2|c_1}(p_1,p_2,p_3)
\cr
+4p_1^{b_3}p_2^{a_3}
\eta^{b_4c_3}T_{2,2,2|3}^{a_1a_2|b_1b_2|c_1c_2}(p_1,p_2,p_3)
+2p_2^{a_3}p_3^{b_3}
\eta^{b_4c_3}T_{2,2,2|2}^{a_1a_2|b_1b_2|c_1c_2}(p_1,p_2,p_3)
\cr
-6ip_3^{a_3}\eta^{b_4c_3}\eta^{a_2b_3}
T_{1,2,2|3}^{a_1|b_1b_2|c_1c_2}(p_1,p_2,p_3)
-2\eta^{b_4c_3}p_3^{a_3}p_1^{b_3}
T_{2,2,2|2}^{a_1a_2|b_1b_2|c_1c_2}(p_1,p_2,p_3)\rbrace
\cr\times
H_{a_1a_2a_3}(p_1)
H_{b_1b_2b_3b_4}(p_2)H_{c_1c_2c_3}(p_3)
}}
It is straightforward to check, by
using the momentum conservation and
by substituting the appropriate expressions for
$T_{p,q,r|s}^{a_1...a_p|b_1...b_q|c_1...c_r}(p_1,p_2,p_3)$
entering (98)
 that the correlator (98)
vanishes identically
on-shell.
In fact, this vanishing is a direct consequence 
of the conformal invariance: comparing the correlators
(97), (98) and, when necessary, using the momentum conservation
${\vec{p}}_1+{\vec{p}}_2+{\vec{p}}_3=0$, 
it is easy to see that each term in (98) has a counterpart in
(97) with precisely the same index structure but with higher
value of $s$ in the appropriate 
$T_{p,q,r|s}^{a_1...a_p|b_1...b_q|c_1...c_r}(p_1,p_2,p_3)$.
For this reason, any term  appearing in (98) is forbidden by the conformal
invariance and has to vanish on-shell (see the discussion above
in this Section).

This concludes the computation of 
all the correlators contributing to the gauge-invariant cubic interaction
of one $s=4$ and two $s=3$ fields.

Summing over all of the contributions
from the three-point correlators (63)-(98), using
the momentum conservation
along with the symmetry of the polarization tensors
 and eliminating terms
that vanish on-shell, we obtain the final answer
for the gauge-invariant 3-point amplitude:
\eqn\grav{\eqalign{<V_{s=3}(p_1)V_{s=4}(p_2)V_{s=3}(p_3)>
\cr
=\lbrace
272\eta^{a_3b_2}\eta^{a_2b_3}\eta^{b_4c_3}
T_{1,1,2|4}^{a_1|b_1|c_1c_2}(p_1,p_2,p_3)
\cr
+144\eta^{a_3b_2}\eta^{b_3c_2}\eta^{b_4c_3}
T_{2,1,1|2}^{a_1a_2|b_1|c_1}(p_1,p_2,p_3)
\cr
-128\eta^{a_2b_3}\eta^{a_3c_2}\eta^{b_4c_3}
T_{1,2,1|2}^{a_1|b_1b_2|c_1}(p_1,p_2,p_3)
\cr
-(16ip_2^{a_3}\eta^{b_3c_2}\eta^{b_4c_3}
+24ip_2^{b_3}\eta^{2_3c_2}\eta^{b_4c_3})
T_{2,2,1|2}^{a_1a_2|b_1b_2|c_1}(p_1,p_2,p_3)
\cr
-32ip_1^{b_3}\eta^{a_3b_2}\eta^{b_4c_3}
T_{2,1,2|4}^{a_1a_2|b_1|c_1c_2}(p_1,p_2,p_3)
+(48ip_1^{c_3}\eta^{a_3b_4}\eta^{a_2b_3}
\cr
+72ip_1^{b_3}\eta^{a_2b_4}\eta^{a_3c_3}
-144ip_2^{a_3}\eta^{a_2b_3}\eta^{b_4c_3})
T_{1,2,2|4}^{a_1|b_1b_2|c_1c_2}(p_1,p_2,p_3)
\cr
+((56-20({\vec{p}}_1{\vec{p}}_2))\eta^{a_3b_3}\eta^{b_4c_3}
-24p_3^{b_3}p_3^{a_3}
\cr
-8p_1^{b_3}p_1^{b_4}\eta^{a_3c_3}-20
p_1^{b_3}p_1^{c_3}\eta^{a_3b_4})
T_{2,2,2|4}^{a_1a_2|b_1b_2|c_1c_2}(p_1,p_2,p_3)\rbrace
\cr\times{I({\vec{p}}_1{\vec{p}}_2)}
H_{a_1a_2a_3}(p_1)
H_{b_1b_2b_3b_4}(p_2)H_{c_1c_2c_3}(p_3)\delta({{p}}_1+{{p}}_2+{{p}}_3)
\cr
+
{\lbrace}
24\eta^{a_2b_4}\eta^{a_3\lbrack{b_3}}\eta^{b_5\rbrack{c_3}}
T_{1,3,2|4}^{a_1|b_1b_2b_3|c_1c_2}(p_1,p_2,p_3)
\cr
+8\eta^{b_4c_3}(ip_1^{b_3}\eta^{a_3b_5}-ip_1^{b_5}\eta^{a_3b_3})
T_{2,3,2|4}^{a_1a_2|b_1b_2b_3|c_1c_2}(p_1,p_2,p_3)
\rbrace\cr\times
{I({\vec{p}}_1{\vec{p}}_2)}
H_{a_1a_2a_3}(p_1)
H_{b_1b_2b_4b_5}(p_2)H_{c_1c_2c_3}(p_3)\delta({{p}}_1+{{p}}_2+{{p}}_3)
\cr
+{\lbrace}
24\eta^{a_3b_3}\eta^{b_4\lbrack{a_4}}\eta^{a_1\rbrack{c_3}}
T_{2,2,2|4}^{a_1a_2|b_1b_2|c_1c_2}(p_1,p_2,p_3)
\cr
+16\eta^{a_3b_3}\eta^{b_4\lbrack{a_4}}\eta^{a_1\rbrack{c_3}}\eta^{b_3c_2}
T_{3,2,1|2}^{a_1a_2a_3|b_1b_2|c_1}(p_1,p_2,p_3)\rbrace
\cr\times
{I({\vec{p}}_1{\vec{p}}_2)}
H_{a_2a_3a_4}(p_1)
H_{b_1b_2b_3b_4}(p_2)H_{c_1c_2c_3}(p_3)\delta({{p}}_1+{{p}}_2+{{p}}_3)\cr
+
\eta^{b_5c_3}\eta^{b_3\lbrack{a_3}}\eta^{a_4\rbrack{b_4}}
T_{3,3,2|4}^{a_1a_2a_3|b_1b_2b_3|c_1c_2}(p_1,p_2,p_3)
\cr\times
{I({\vec{p}}_1{\vec{p}}_2)}
H_{a_1a_2a_4}(p_1)
H_{b_1b_2b_4b_5}(p_2)H_{c_1c_2c_3}(p_3)
\delta({{p}}_1+{{p}}_2+{{p}}_3)}}
The gauge-invariant cubic interaction term in the position
space is then easy to obtain from (99)
by usual Fourier transform.

\centerline{\bf 7. Conclusion and Discussion}

In this paper we have constructed vertex operators
in open string theory, describing massless 
higher spin fields and computed the three-point
correlation function describing the gauge-invariant cubic
interaction of  $s=4$ with two $s=3$ particles.
The computation performed in this paper is straightforward
to generalize to obtain the gauge-invariant cubic interactions
of other massless higher spins, although technically
in certain cases practical computations could be quite complicated
due to the picture changing issue.
The BRST-invariance conditions for the
 vertex operators, constructed in this paper, lead
to Fierz-Pauli on-shell conditions 
for the space-time higher spin fields.
The BRST-nontriviality constraints on these operators lead to the
gauge transformations for the space-time fields:
as the gauge transformations imply shifting  higher
operators by BRST-trivial terms, the correlators are
automatically invariant under these transformations and so are
the interaction terms induced by these correlators.
As we have pointed out, the gauge transformations for spin
$s$ fields, implied by the BRST non-triviality of their vertex operators,
are equivalent to the transformations given by the
symmetrized derivatives of spin $s-1$ gauge parameter, restricted
by the tracelessness constraints. So
the vertex operators, considered in this paper, give a description of 
interacting higher spins, isomorphic to Fronsdal's framework ~{\fronsdal}
rather than a well-known alternative approach 
involving non-local compensators, traded for the tracelessness
constraints on the gauge symmetries {\sorokin, \sagnottie, \sagnottif,
\sagnottig}
It would be interesting to try to interpret the compensator
approach in the language of string theory.
Interestingly, the nonlocality of the massless higher spin interactions,
observed in this paper (already in the cubic case), is the direct
consequence of the non-standard ghost structure of the vertex operators,
leading to the deformation of the  usual Koba-Nielsen's measure
and the appearance of the integrated vertices in three-point
amplitudes - while typically,
as far as the standard lower
spin cases are concerned (such as photon, graviton, etc.) 
three-point amplitudes
only involve unintegrated vertices
and thus no nonlocalities. It should be
noted, however, that in the massive case
one shouldn't expect nonlocalities for higher spins
on the cubic level either, as
the massive higher spin vertex operators appear naturally in the massive
sector of string theory and have standard ghost structures,
not different from the lower spin case. Thus the nonlocalities
in cubic interactions, observed in this work, appear to be specific 
to the massless case only.

In this work we have considered, for simplicity,
 the higher spin vertex operators 
for the values of $s$ from 3 to 9 in
the totally symmetric case. It should be, however, 
quite a direct excersise to extend our calculation
to less symmetric cases, including those involving
several families of indices, although matching the gauge
symmetries of vertex operators on the string theory side
(as a consequence of the BRST conditions on the operators)
to the  standard gauge symmetries
observed in higher spin field theories in space-time,
 is an open question. As we have pointed out, 
this matching does work out in totally symmetric case,
considered in this paper; in less symmetric cases this 
will require additional careful analysis of the BRST constraints
on the appropriate vertex operators. We hope to perform this analysis
in our future work.
Another important direction for the future work is to extend
the formalism developed in this work to compute the 
higher order
gauge-invariant interaction terms of the higher spin fields, such 
as the quartic interactions. As in the cubic case, one would expect
the non-localities, stemming from the ghost structures 
of the vertex operators; in addition the calculation of the four-point
functions will require careful analysis of ghost number balance and 
(whenever necessary) insertions of appropriate picture changing operators.
It is also possible that the ghost number balance conditions shall impose
certain selection rules  for the higher order interactions.
In general though, it appears that string theory provides us
with a set of  powerful tools to investigate the interacting theories
of the higher spin fields. What seems especially attractive about the
 string-theoretic approach, is that the traditionally difficult
issues about the higher spin field theories (such as the gauge invariance
of the interaction terms )
appear to be under control
in string theory - e.g. with the BRST conditions automatically
ensuring the the 
gauge invariance in the space-time effective action.
In this work we have restricted ourselves to the totally symmetric 
higher spin fields, appearing in open string theory framework.
 Considering higher spins related to less trivial
Young tableau and with several families of indices will 
particularly require
to extend the analysis and the formalism, developed in this paper, to the 
closed string case. This is another direction for the future
research and the subject for the work currently in progress.

\centerline{\bf Acknowledgements}

I would like to thank  
Massimo Bianchi, Robert De Mello Koch, 
 and Augusto Sagnotti for useful comments and discussions.
I'm also grateful to Evgeniy Skvortsov for
very useful comments on 
 the gauge symmetry transformations
for the higher spin fields,
as well as for the pointing
out some typos in the previous version of this paper.

\listrefs

\end